\documentclass{aa}

\usepackage[varg]{txfonts}
\usepackage{graphicx}
\usepackage{booktabs}

\usepackage{amsmath}	
\usepackage{amssymb}	
\usepackage{xcolor}
\usepackage{siunitx}
\usepackage{makecell}

\usepackage[]{hyperref}
\hypersetup{colorlinks,citecolor=blue, linkcolor=blue, urlcolor=blue, breaklinks=true}

\newcommand{\micron}{$\mu m$}

\begin{document}

\title{A quiescent galaxy in a gas-rich cosmic web node at z$\sim$3}

\titlerunning{A quiescent galaxy in a gas-rich cosmic web node at z$\sim$3}
\authorrunning{Weichen Wang et al.}

\author{Weichen Wang\inst{\ref{unimib}}\fnmsep\thanks{\email{weichen.wang@unimib.it}}
\and 
Sebastiano Cantalupo\inst{\ref{unimib}}
\and 
Marta Galbiati\inst{\ref{unimib}}
\and 
Andrea Travascio\inst{\ref{unimib},\ref{trieste}}
\and 
Antonio Pensabene\inst{\ref{unimib},\ref{dawn},\ref{dtu}}
\and 
Charles C. Steidel\inst{\ref{caltech}}
\and 
Gabriele Pezzulli\inst{\ref{groningen}}
\and 
Bingjie Wang\inst{\ref{princeton}\fnmsep\thanks{NHFP Hubble Fellow}}
\and 
Xiaohan Wang\inst{\ref{tsinghua},\ref{unimib}}
\and 
Rajeshwari Dutta\inst{\ref{iucaa}}
\and 
Titouan Lazeyras\inst{\ref{unimib}}
\and 
Nicolas Ledos\inst{\ref{unimib}}
\and 
Huiyang Mao\inst{\ref{pmo},\ref{unimib}}
\and 
Giada Quadri\inst{\ref{unimib}}
}

\institute{Department of Physics, Universita degli Studi di Milano-Bicocca, I-20126 Milano, Italy\label{unimib}
\and 
INAF – Osservatorio Astronomico di Trieste, I-34131 Trieste, Italy \label{trieste}
\and 
Cosmic Dawn Center (DAWN), Copenhagen, Denmark \label{dawn}
\and 
DTU Space, Technical University of Denmark, DK-2800 Kgs. Lyngby, Denmark \label{dtu}
\and
Cahill Center for Astronomy and Astrophysics, California Institute of Technology, Pasadena, CA 91125, USA\label{caltech}
\and 
Kapteyn Astronomical Institute, University of Groningen, NL-9747 AD Groningen, the Netherlands \label{groningen}
\and 
Department of Astrophysical Sciences, Princeton University, Princeton, NJ 08544, USA\label{princeton}
\and 
Department of Astronomy, Tsinghua University, Beijing 100084, China\label{tsinghua} 
\and 
IUCAA, Postbag 4, Ganeshkind, Pune 411007, Maharashtra, India \label{iucaa}
\and 
Purple Mountain Observatory, Chinese Academy of Sciences, Nanjing
210023, China \label{pmo}
}

\date{\today}

\abstract{
Recent JWST observations have unveiled a large number of quiescent galaxies at $z\gtrsim 3$, bringing potential challenges to current galaxy formation models. Since star formation is expected to be fed by external gas accretion, knowledge about the circumgalactic media (CGM) of these galaxies is essential to understanding how they quench. In this work, we present the discovery of a massive, passive galaxy ($M_\star \simeq 10^{11}\,M_\sun$) within the MUSE Quasar Nebula 01 (MQN01) large-scale structure at $z\simeq 3.25$, containing one of the largest overdensities of galaxies and active galactic nuclei (AGNs) found so far at $z\gtrsim 3$. This passive galaxy has a star formation rate of  $4^{+6}_{-2}\,M_\sun/$yr, placing it more than 1~dex below the star-forming main sequence, and it has no detectable molecular gas ($M_\mathrm{H2}<2\times 10^{10}~M_\sun$ within a 700\,km s$^{-1}$ window, if assuming $r_{41}=0.34$ and $\alpha_\mathrm{CO} = 4\ M_\sun~\mathrm{K^{-1}~km^{-1}~s~pc^{2}}$). Surprisingly, it is located at the center of a large cool gas reservoir, as traced by bright Ly$\alpha$ and H$\alpha$ emission. 
Deep multiwavelength information unique to this field, including spectroscopic and \emph{Chandra} X-ray observations, suggests that an
external mechanism could keep this cool gas reservoir in a dynamically hot, turbulent state, thus reducing its ability to efficiently accumulate within the galaxy.
In particular, we discuss the possibility that the agent of this external energy injection is the AGN jet from a nearby star-forming galaxy located at a projected distance of 48 kpc and detected in both X-ray and radio. Additional or alternative energy injection mechanisms could involve the gravitational interaction between these two galaxies.
In all scenarios, the elevated ionizing field provided by the AGN overdensity, including the nearby AGN, can illuminate the passive galaxy's cool CGM and make it visible through fluorescent emission. 
Our study demonstrates that the star formation rates of high-redshift galaxies could be substantially reduced and maintained at a low level even within gas-rich and overdense environments in particular situations.
}


\maketitle
\nolinenumbers

\section{Introduction} \label{sec:intro}

The formation of galaxies is connected to the large-scale cosmic environments in which they are located. To put it in a simplified way, galaxies commonly accrete gas from the cosmic web, convert it to stars, and grow in stellar mass since the very early phases of their formation (\citealt{Blumenthal1984,Keres2005,Tacconi2020}). As such, two key elements of galaxy formation become closely connected to each other, namely the gas in the galaxies' surroundings and the star formation. This whole process, or any individual step of it, may also be slowed down or halted, leading to the emergence of quiescent galaxies (\citealt{Faber2007,Man2018,Afruni2019}).

The process of converting the gas into the stars appears to be more efficient for galaxies at high redshifts (\citealt{Noeske2007,Madau2014,ForsterSchreiber2020}) and overdense environments (\citealt{Steidel2005,Elbaz2007,Chiang2017}). 
Specifically, studies of galaxy protoclusters or cosmic web nodes at $z\gtrsim 3$ reported that these environments both host rich reservoirs of cool gas ($10^{4}-10^5$~K), which is commonly traced by Ly$\alpha$ emission, and molecular gas (e.g., \citealt{Steidel2000,Borisova2016,Cantalupo2017,Pensabene2024,Perez-Martinez2025,Zhou2025}). They are different from their local descendants hosting mainly hot gas ($10^{7}-10^8$~K; \citealt{Alberts2022}). Overdense environments at early epochs also see an excess of massive star-forming galaxies with respect to the field (e.g., \citealt{Elbaz2007,Koyama2013,Galbiati2025,Zhou2025a,GWang2025}).  Therefore, the $z\gtrsim 3$ cosmic web nodes are expected to be where the formation of massive galaxies via gas accretion is exceptionally efficient.

From this perspective, gas-rich cosmic web nodes are uniquely valuable laboratories for investigating which physical processes are most important in facilitating early-phase galaxy formation. As an example study of this kind, a recent work by \cite{Wang2025} reports a giant massive star-forming disk forming in such a cosmic web node, suggesting that corotating gas inflows could play an important role in the formation of these star-forming systems (see also \citealt{Venkateshwaran2024,Jiang2025,Umehata2025a,Pensabene2025}). 

The emergence of passive galaxies in protocluster environments has not been studied as thoroughly thus far.
Recent observations utilizing the James Webb Space Telescope (JWST) or ground-based  facilities have made remarkable progress in searching for high-redshift quiescent galaxies in the general field population (e.g., \citealt{Ilbert2013,Glazebrook2017,Glazebrook2024,DEugenio2020,Valentino2020,Santini2021,Nanayakkara2022,Carnall2023,DEugenio2024,Baker2025,deGraaff2025,delaVega2025,Weibel2025}). These studies find abundant massive quiescent systems at $z\gtrsim3$, with number densities exceeding the predictions from some models or simulations (\citealt{Valentino2023,Carnall2024,Lagos2025}). However, the large-scale environments of these systems remain poorly explored. A few studies suggest that they live in overdense environments based on the number counts of neighbor galaxies (\citealt{Kubo2021,Kubo2022,Ito2023,Carnall2024,Jin2024,Kakimoto2024,Tanaka2024,Alberts2024,Forrest2024,deGraaff2025,McConachie2025,Witten2025}). On the other hand, observational probes of their circumgalactic medium (CGM) are still scarce \citep{Kalita2021,Umehata2025b,PerezGonzalez2025,Guo2025}, which, however, are essential to understanding the regulation of galaxy growth and quenching at high redshifts. 

In this work, we present the discovery of a massive quiescent galaxy in a gas-rich environment of a cosmic web node or protocluster at $z\sim3.2$, identified and spectroscopically confirmed from a JWST program. Almost unique with respect to other systems found so far, the cool component of the CGM of this galaxy is visible through fluorescent Ly$\alpha$ emission due to the presence of a large overdensity of AGNs in the field (\citealt{Cantalupo2005,Cantalupo2014,Borisova2016,Travascio2025}). Moreover, deep multiwavelength datasets are available for this field, including JWST, Hubble Space Telescope (HST), Very Large Telescope (VLT), Atacama Large Millimeter/submillimeter Array (ALMA), and \emph{Chandra}, making it possible to characterize the detailed properties of the gas in and around galaxies. Other identified galaxies in the structure that are discovered from the Near Infrared Spectrograph (NIRSpec) observations of the JWST program are all star-forming and are presented in a separate paper (\citealt{XWang2025}). In addition, a broader study of photometrically selected quiescent member galaxies will be  presented in a companion paper (M.~Galbiati et al.~in prep.).

The observations and data reduction are described in Sect.\ \ref{sec:data}. Measurements of the physical properties of the galaxy are presented in Sect.\ \ref{sec:measurements}, with further technical details provided in the appendices. An analysis of the environment of the galaxy is presented in Sect.\ \ref{sec:measurements_env}.  Discussions about the implications of the observational results are presented in Sect.\ \ref{sec:discussion}. Finally, a short summary is provided in Sect.\ \ref{sec:summary}. A lambda cold dark matter ($\Lambda$CDM) cosmology with $H_0 = 70~$km/s/Mpc, $\Omega_m = 0.3$, and $\Omega_\Lambda = 0.7$ is adopted throughout the paper.

\section{Observations and data reduction} \label{sec:data}

\begin{figure*}
\centering
\includegraphics[width = 5.7in]{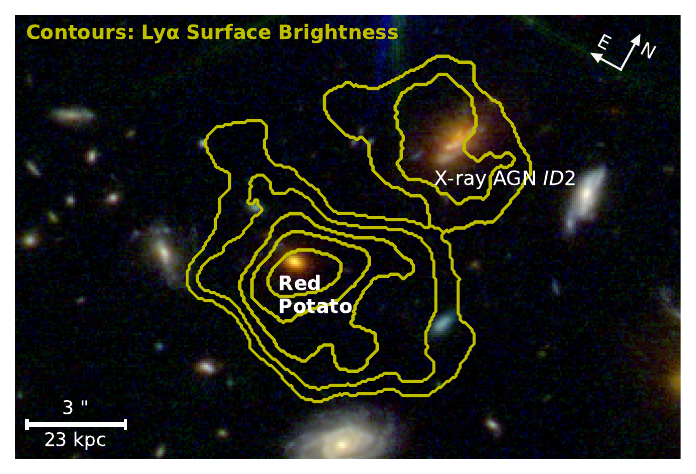}
\caption{Passive galaxy MQN01~J004131.9-493704 ``Red Potato'' at $z=3.250$ and its surrounding cool Ly$\alpha$-emitting gas reservoir. 
This massive and compact object was discovered at the center of an extended $10^4$--$10^5$~K gas reservoir, as traced by the bright Ly$\alpha$ emission spanning 10 arcsec (contours) within an overdense region \citep{Pensabene2024,Galbiati2025,Travascio2025}. 
An X-ray bright AGN-\emph{ID2} from \citet{Travascio2025} was found at a projected distance of only 7 arcsec or 50 kpc and a line-of-sight velocity separation of 100 km/s, likely providing enough ionizing photons to illuminate the CGM of the Red Potato and the CGM of its host galaxy. The color image was made using the NIRCam/F322W2, NIRCam/F150W2, and ACS/F814W for the red, green, and blue channels. The Ly$\alpha$ surface brightness values are 10, 7.0, 5.5, 4.0, 3.0 ($\times 10^{-18}~$erg/s/cm$^2$/arcsec$^2$) from the outermost to the innermost contour around the Red Potato, and 4.0, 3.0 for the contours around the AGN-\emph{ID2}. \label{fig:rgb_w_lya} }
\end{figure*} 

\subsection{Near-infrared and optical imaging}
\label{subsec:imaging}

The target galaxy of this work is named MQN01~J004131.9-493704, which we also call the ``Red Potato.'' It was observed with the JWST Near Infrared Camera (NIRCam) as part of the program GO1835 (F150W2 \& F322W2, 1632s; P.I. Cantalupo). In addition, it was also observed with HST as parts of the programs ID 17065 using the Advanced Camera for Surveys (ACS) in the F625W and F814W bands (10 \& 12 orbits; P.I. Cantalupo) and the program ID 17483 using the Wide Field Camera 3 (WFC3) in the F160W band (1 orbit; P.I. Dutta). Details of the observations and reductions of the NIRCam and ACS images can be found in \cite{Wang2025}. The WFC3 image was reduced with the pipeline \textsc{calwf3}\footnote{https://wfc3tools.readthedocs.io/en/latest/wfc3tools/calwf3.html}. Additional observations with the VLT High Acuity Wide field K-band Imager (HAWKI)
 instrument were conducted in three filters, H, Ks, and the CH4 1.58\micron\ narrow filter, as part of the program ID 110.23ZX (P.I. Cantalupo). The exposure time was 1-2 hr for each filter. Details of the data reduction can be found in \cite{Wang2025} and \cite{Galbiati2025}. All images were aligned to \emph{Gaia} astrometry using {\sc tweakwcs}\footnote{https://tweakwcs.readthedocs.io/} before being measured for fluxes with {\sc Source Extractor} (v.~2.28; \citealt{1996A&AS..117..393B}). The point spread function (PSF) profiles for these image filters were made from isolated stars using {\sc photutils} \citep{Bradley2024}.

To prepare for flux measurements in the JWST and HST filters, the images were resampled to the same pixel grid using {\sc drizzlepac} \citep{Fruchter2002} and PSF-matched using {\sc photutils}. The Source Extractor runs were done in dual mode, using the F322W2 as the detection band as described in \cite{Wang2025}. Flux uncertainties were inferred by placing and measuring flux counts from empty sky apertures with the same area as that used for the galaxy photometry. The images in the HAWKI filters were processed and analyzed in the same way as described above, except that the detection band was made by coadding the H and Ks images. 

A total of eight photometric bands are available and included for analysis of this work: three HST filters, ACS/F625W, ACS/F814W, and WFC3/F160W; two JWST/NIRCam filters, F150W2 and F322W2; three VLT/HAWKI filters, CH4, H, and Ks. Along with the ALMA band 6 at observed 1.2~mm (Sect.\ \ref{subsec:xray_radio_submm}),
these filters capture the rest-frame ultraviolet (UV) to near-infrared (IR) part of the galaxy spectral energy distribution (SED) and the potential dust continuum emission.

\begin{table*}
\caption{Properties of the Red Potato galaxy MQN01~J004131.9-493704.}\label{tab:table1}
\centering
\begin{tabular}{@{}llll@{}}
\hline 
Property & Value & Unit & Reference\\
\hline
Right ascension (J2000) & 00h41m31.88s & - & - \\
Declination (J2000) & -49d37m03.7s & - & - \\
Redshift & 3.2496$\pm$0.0006 & - & Sect.\ \ref{subsec:nirspec_muse} \\
Stellar mass & $1.1^{+0.4}_{-0.4}$ & $10^{11} M_\odot$ & Sect.\ \ref{subsec:analysis_sedfitting_sfr_agn} \\
SFR (SED fitting) & $4^{+6}_{-2}$ & $M_\odot$/yr & Sect.\ \ref{subsec:analysis_sedfitting_sfr_agn}  \\
SFR (H$\alpha$) & $<16$ & $M_\odot$/yr & Sect.\ \ref{subsec:analysis_sedfitting_sfr_agn}  \\
SFR (UV+IR) & $<19$ ~if adopting SMG templates & $M_\odot$/yr & Sect.\ \ref{subsec:analysis_sedfitting_sfr_agn}  \\
Molecular gas mass ($M_\mathrm{H_2}$) & $<2.0 \cdot (\alpha_\mathrm{CO}/4~M_\sun~\mathrm{K^{-1}~km^{-1}~s~pc^{2}})$ & $10^{10} M_\odot$ & Sect.\ \ref{subsec:analysis_fgas} \\
Molecular gas fraction ($f_\mathrm{H2}$) & $<0.15 \cdot (\alpha_\mathrm{CO}/4)$ & - & Sect.\ \ref{subsec:analysis_fgas} \\
Stellar velocity dispersion (integrated) & 268$\pm$20 & km/s & Sect.\ \ref{subsec:analysis_kinematics_outflows} \\
Half-light radius (along major axis) & 1.0 (1.5\,\micron), 0.8 (3.2\,\micron) & kpc & Sect.\ \ref{subsec:analysis_morphology} \\
\hline 
\end{tabular}
\end{table*}

\subsection{Near-infrared and optical spectroscopy}
\label{subsec:nirspec_muse}

The Red Potato galaxy MQN01~J004131.9-493704 was observed with the JWST NIRSpec Micro Shutter Array (MSA) as part of the program GO 1835. A filter and grating combination of F170LP/G235H was used, leading to a spectroscopic resolution of $R= 2000$-$3700$ in the wavelength range of 1.7-3.2~$\mu m$. The total exposure time on source was 7 hr. More details about the observation setup can be found in \cite{Wang2025}. 

Data reduction was conducted using the official {\sc jwst} pipeline version 1.16.0 and the Calibration Reference Data System version jwst\_1290.pmap. The Red Potato galaxy shows a bright continuum on the 2D spectrum, the spatial profile of which can be fit with a Gaussian with a width of 1.1 NIRSpec pixel (0.12 arcsec) in standard deviation. A 3-$\sigma$ spatial window, corresponding to a full width of 7 pixels, was adopted for the galaxy 1D spectrum extraction. The background was measured locally from sky regions in the same slit which are 11-16 pixels away from the galaxy center, and subtracted from the galaxy spectrum. A flux calibration was performed onto the extracted spectrum using the integrated galaxy fluxes measured from the imaging data, in order to correct for the flux loss out of the NIRSpec slit, which is described in Appendix \ref{sec:appendix_slitloss}.

The galaxy redshift is 3.2496$\pm$0.0006, which was determined by fitting the galaxy stellar continuum using \textsc{ppxf} (\citealt{Cappellari2023}; Sect.\ \ref{subsec:analysis_kinematics_outflows}). The uncertainty quoted for the redshift is at 2-$\sigma$, corresponding to a velocity uncertainty of 40 km/s, and it was calculated by bootstrapping the residual noise in the fitting following \cite{Cappellari2023}.

The galaxy and its surrounding field were also observed by the VLT Multi-Unit Spectroscopic Explorer (MUSE) with a depth of 10 hours, as part of a Guaranteed Time Observations Program (ID 0102.A-0448; P.I. Cantalupo). Details regarding the observations, data reduction, postprocessing, extraction of the Ly$\alpha$ line emission  will be presented in Cantalupo et al. (in prep.; see also \citealt{Galbiati2025}). The galaxy is detected with spatially extended Ly$\alpha$ emission, the radial extent of which will be presented in Sect.\ \ref{sec:measurements}.

\begin{figure*}
\centering
\includegraphics[width = 6.8in]{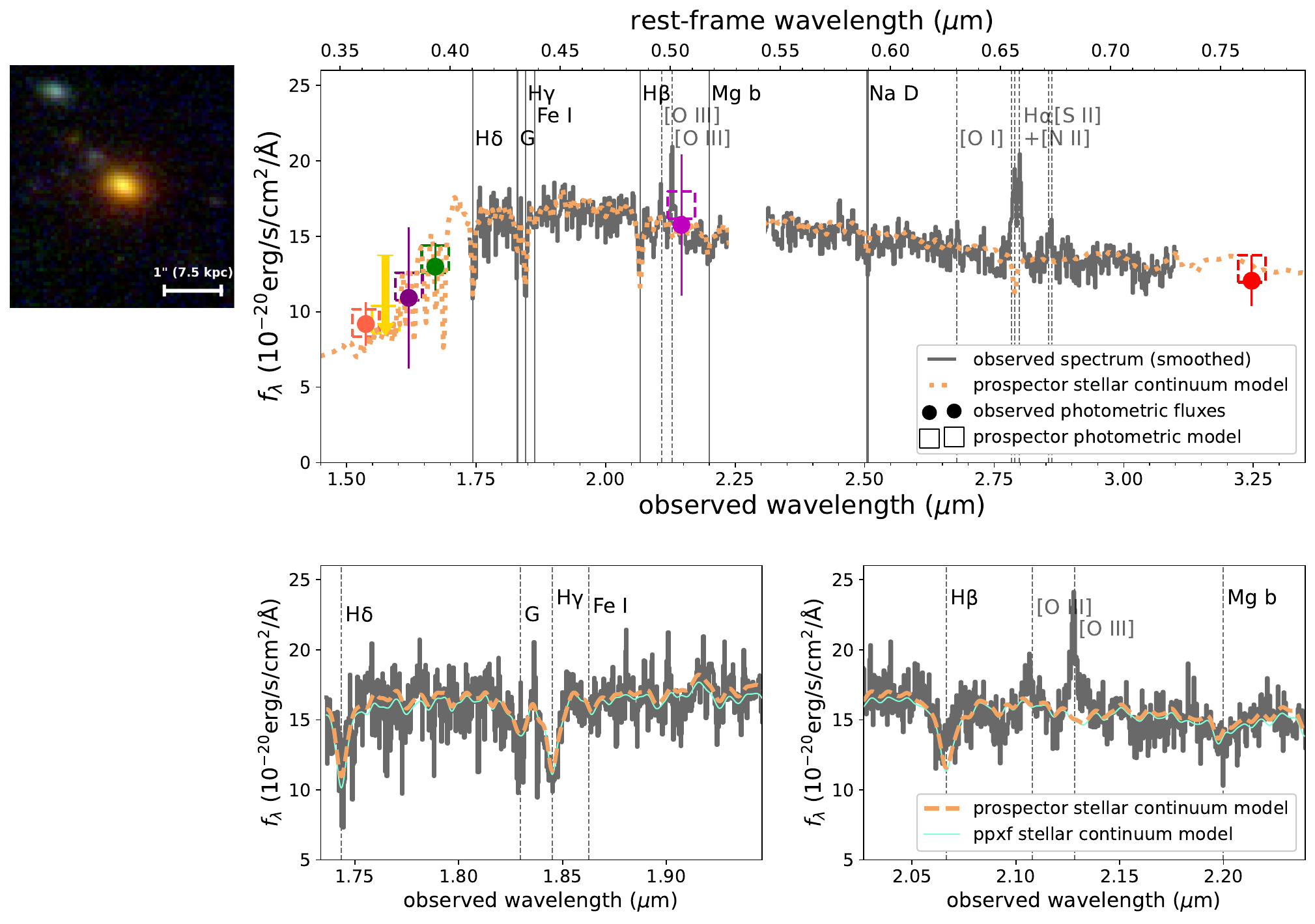}
\caption{Spectrum and multiband photometry of the Red Potato galaxy suggesting its quiescent nature. \emph{Top:} The image of the Red Potato galaxy is shown on the left. Its NIRSpec spectrum, which is smoothed by a 5-pixel box kernel, shows strong Balmer and metal absorption lines (solid vertical lines), suggesting old stellar populations. This is also supported by the D4000 break captured by the multiband photometry (filled circles). Best-fit models of the stellar continuum and photometric fluxes from {\sc prospector} are shown as the orange dotted line and open squares, respectively. Nebular emission lines are also present in the observed spectrum, which are however driven by an AGN-like radiation field (Fig.~\ref{fig:nirspec_zoomins} and Sect.\ \ref{subsec:discussion_quenching_galaxy}) rather than internal star formation. A 2-$\sigma$ errorbar is also plotted for each photometric point, except for the CH4 band where a 2-$\sigma$ upper limit is plotted (yellow arrow). \emph{Bottom:} Zoomed-in views of the absorption lines, including H$\delta$, G, H$\gamma$, Fe~I, H$\beta$ (blended with a weak emission component), and Mg~b.
\label{fig:nirspec_w_models} }
\end{figure*}

\subsection{X-ray, submillimeter, and radio  observations}
\label{subsec:xray_radio_submm}

The Red Potato galaxy was observed with the \emph{Chandra} Advanced CCD Imaging Spectrometer (Cycle 23, 634 ks; P.I. Cantalupo) and ALMA (Program ID 2021.1.00793.S, 20 hr; P.I. Cantalupo). Details of the observations and data reduction can be found in \cite{Travascio2025} and \cite{Pensabene2024}, respectively. No significant X-ray detection is found at the location of the galaxy, corresponding to 2-$\sigma$ upper limits of $1.7\times 10^{-16}~$erg/s/cm$^2$ and $2.9\times 10^{-16}~$erg/s/cm$^2$ for 0.5-2 keV and 2-7 keV, respectively \citep{Travascio2025}. Regarding the ALMA observations, the galaxy is not detected with either the 1.2\,mm continuum in Band 6 or the CO (4-3) line in Band 3, corresponding to 5-$\sigma$ limits of 0.2 mJy and 4.2$\times10^{9}\ \mathrm{K~km~s^{-1}~pc^2}$, respectively, for a velocity window of 700 km/s ($\simeq 2\times$\,integrated stellar velocity dispersion) and the beam size of 1.4\arcsec\ \citep{Pensabene2024}.

The galaxy was also observed in radio at 0.8 GHz and 1.4 GHz by the Rapid ASKAP Continuum Survey (RACS; \citealt{McConnell2020,Hale2021,Duchesne2024}). The radio images and source flux catalogs were obtained from the ASKAP Science Data Archive\footnote{https://research.csiro.au/casda/}. The RACS maps are known to be subject to astrometric uncertainties of a few arcseconds (\citealt{McConnell2020,Duchesne2024}). Thus, we conducted astrometry alignment of the radio images using the bright quasi stellar object (QSO) of MQN01, CTS G18.01. Specifically, we conducted the alignment by matching the QSO peak locations measured on the ASKAP images to the QSO position measured on the JWST NIRCam image. This led to an astrometry correction of $\Delta \alpha \cdot \cos \delta = -1.9$", $\Delta \delta = -1.0$" for the 0.8 GHz image and a correction of $\Delta \alpha \cdot \cos \delta = -1.6$", $\Delta \delta = 0.0$" for the 1.4 GHz image, which are within the ranges of astrometry errors reported in the RACS survey papers (\citealt{McConnell2020,Duchesne2024}). The aligned radio maps in both bands can be found in Appendix \ref{sec:appendix_askap}. Details about further analysis of the radio data are deferred to Sect.\ \ref{subsec:analysis_xray_radio}.

\section{Analysis of galaxy properties} 
\label{sec:measurements}

The Red Potato galaxy MQN01~J004131.9-493704 is a massive red system at $z=3.250$ discovered at the center of an extended cool ($10^4$-$10^5$~K) gas reservoir traced by the Ly$\alpha$ emission, as shown in Fig.~\ref{fig:rgb_w_lya}. Intriguingly, it has a low star formation rate (SFR), at least 1 dex below the star-forming main sequence (SFMS) at its redshift.
Detailed physical properties of the galaxy are listed in Table~\ref{tab:table1}, whereas in the following we describe details of the relevant measurements.

\begin{figure*}
\centering
\includegraphics[width = 7.2in]{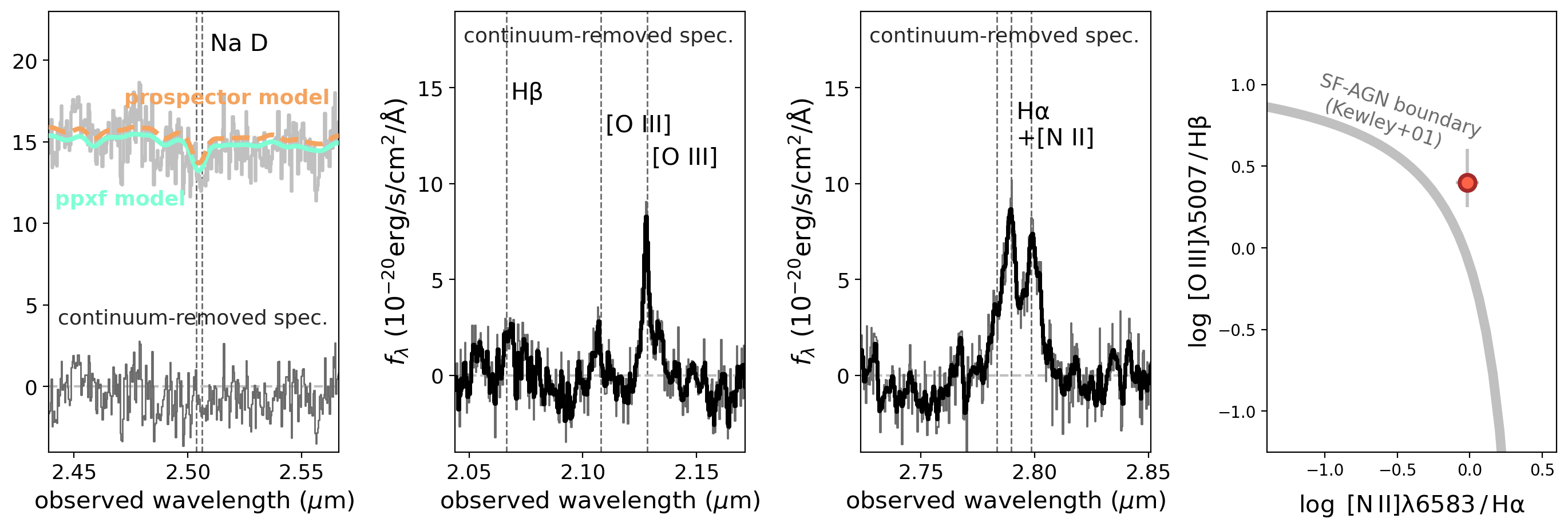}
\caption{Profiles of the Na~D and nebular lines on the continuum-removed spectrum of the galaxy. \emph{Left panel:} The Na D line profiles before (light gray) and after (black) removing the stellar continuum from {\sc prospector} (orange dashed). The continuum-subtracted spectrum shows no residual absorption component from gas, suggesting the absence of cold ISM gas and outflows. The continuum model from \textsc{ppxf} is also shown (cyan solid) which is consistent with the {\sc prospector} model. \emph{Middle panels:} Hydrogen, [O III], and [N II] lines are detected from the continuum-subtracted spectrum. The black curves represent the spectrum smoothed with a boxcar kernel of 4 pixels and the light gray represent the native-resolution spectrum. \emph{Right panel:} The observed emission lines are driven by an AGN-like spectrum rather than star formation, according to the BPT line ratio criteria by \cite{Kewley2001} and are consistent with the presence of the nearby AGN-\emph{ID2} or, in general, with an elevated ionizing field produced by the large AGN overdensity in this region. Ionization by old post-AGB stars can also produce elevated [N II]/H$\alpha$ in elliptical galaxies (\citealt{Binette1994}), but we can exclude it as a dominant mechanism for Red Potato according to its stellar mass and observed H$\alpha$ luminosity.
}

\label{fig:nirspec_zoomins}
\end{figure*}

\subsection{Spectral lines and the full SED}
\label{subsec:analysis_spectrum_sed}

The quiescent nature of the galaxy is first supported by the absorption lines detected in its NIRSpec spectrum and the shape of its full SED. The spectrum and full SED are shown in Fig.~\ref{fig:nirspec_w_models}. 

The stellar absorption lines indicate old stellar populations formed from a prolonged period of star formation. Specifically, strong Balmer absorption lines (H$\beta$, H$\gamma$, H$\delta$) and the metal lines (G band 4304\AA, Mg~b triplet 5175\AA, Na~D 5894\AA) are all present in the NIRSpec spectrum (Fig.~\ref{fig:nirspec_w_models}). The Balmer lines are contributed by stars with typical ages of several hundred Myrs, whereas the G band and Mg b triplet by stars older than 1 Gyr (e.g., \citealt{Vazdekis2010}). These absorption lines are also commonly found in other high-redshift quiescent systems (\citealt{Nanayakkara2022,Glazebrook2024}). 

The SED shows a drop below rest-frame $0.4\,\mu $m according to the broadband photometry, which is plotted as filled circles in Fig.~\ref{fig:nirspec_w_models}. This indicates a significant D4000 break and hence the presence of old stellar populations. The filter names of the photometric measurements can be found later in Fig.~\ref{fig:prospector_full}. 

A few nebular emission lines are also detected in the galaxy spectrum, including the [O III] 4959\AA, [O III] 5007\AA, H$\alpha$, and the [N II] 6548/6583\AA\ doublet, as shown in Fig.~\ref{fig:nirspec_zoomins}. However, an analysis of the line flux ratios, described in the following Sect.\ \ref{subsec:analysis_sedfitting_sfr_agn}, indicates that they are produced by an AGN-like spectrum rather than by 
ongoing star formation.

\subsection{Stellar mass, SFR, and AGN activity}
\label{subsec:analysis_sedfitting_sfr_agn}

\begin{figure*}
\centering 
\includegraphics[width = 6.8in]{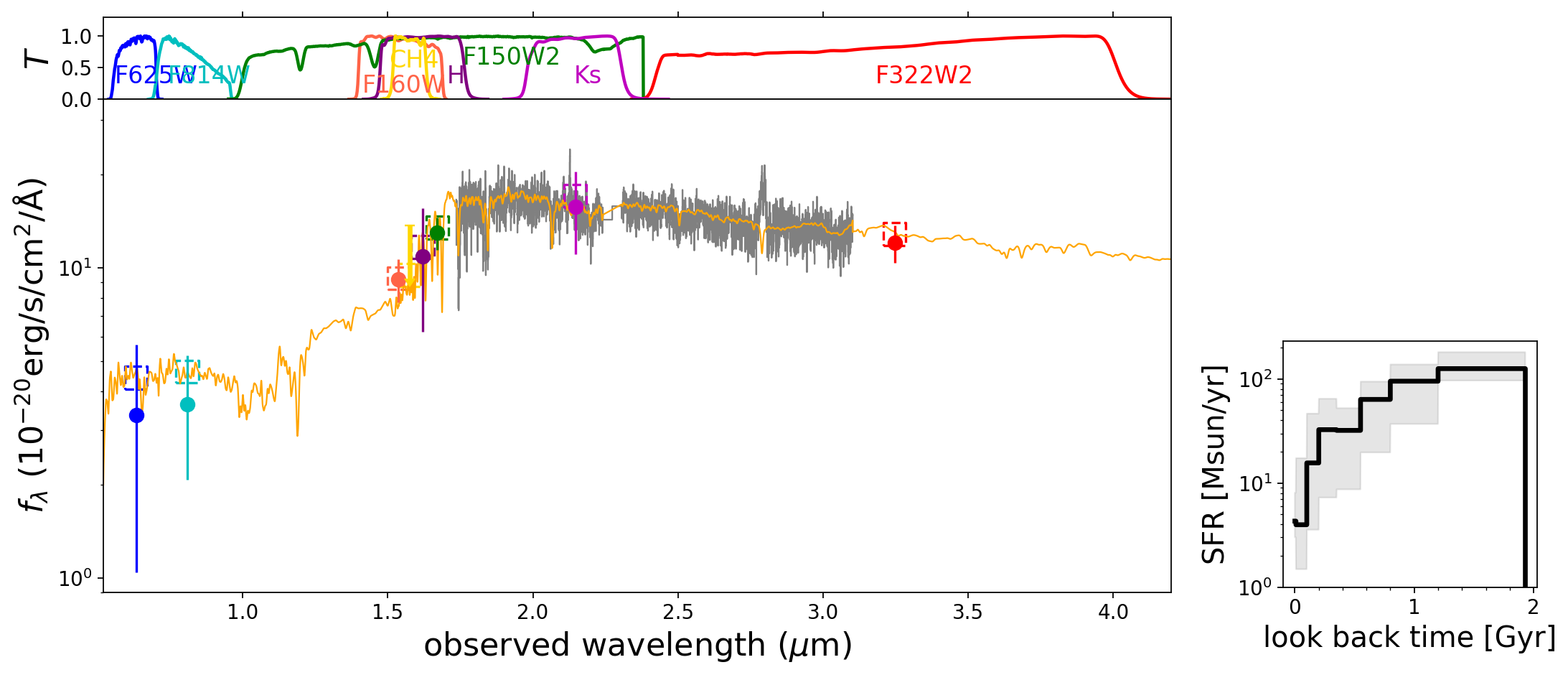}
\caption{Full galaxy SED measured from the NIRSpec spectroscopy (gray curve) and multiband photometry (filled circles), and corresponding models from {\sc prospector} (orange curve and open squares). The {\sc prospector} fitting results suggest a SFR below 10 M$_\sun$/yr, demonstrated by the best-fit SFH in the right panel, corresponding to a specific SFR $< 10^{-10} \mathrm{yr}^{-1}$. The low SFR is also supported by the faint rest-frame UV fluxes (in F625W and F814W filters) and nondetection of dust continuum emission from ALMA (not shown in figure). 
During the {\sc prospector} fitting, the observed emission lines were neglected in the step of modeling the stellar population and SFH, considering that these lines are likely not produced by star formation (Fig.~\ref{fig:nirspec_zoomins}).
The transmission curve of each filter is shown at the top.  \label{fig:prospector_full} }
\end{figure*}

The  galaxy is a massive quiescent system at least one order of magnitude below the SFMS at its epoch, according to its stellar mass and SFR values inferred from the observed photometric fluxes and spectrum. The fluxes and spectrum, along with a best-fit model spectrum obtained using the {\sc prospector} code \citep{Leja2017,Johnson2021,Wang2024}, are presented in Figs.~\ref{fig:nirspec_w_models} \& ~\ref{fig:prospector_full}. 

The {\sc prospector} fitting was done simultaneously onto the observed photometric fluxes in eight filters from optical to near-infrared (Sect.\ \ref{subsec:imaging}) and the NIRSpec spectrum. A nonparametric star formation history (SFH) was adopted, with a prior favoring constant star formation and continuity as described in \cite{Leja2019} and eight lookback time bins of 0-10 Myr, 10-100 Myr, 100-200 Myr, 200-350 Myr, 350-550 Myr, 550-800 Myr, 800-1200 Myr, and 1200-1930 Myr. A \cite{Chabrier2003} initial mass function and a flexible dust attenuation law by \cite{Charlot2000} were adopted. The stellar metallicity was set as a free parameter with a flat prior. The Balmer, [N II], and [O III] emission lines were fit with standalone Gaussians and their contributions to the broadband fluxes are taken into account in the SED fitting. However, because these lines are driven by AGNs as shown below, they were neglected for the step of modeling the SFH and stellar populations in the {\sc prospector} fitting (see also, for example, \citealt{deGraaff2025} for a similar treatment). 
The model spectrum was smoothed by the instrument line spread function and a Gaussian with the same width as the measured stellar velocity dispersion (Sect.\ \ref{subsec:analysis_kinematics_outflows}) before being fit to the NIRSpec data. Following the practices adopted by studies involving fitting the galaxy NIRSpec spectra with {\sc prospector} (e.g., \citealt{Beverage2025,deGraaff2025}) and other tools (e.g., \citealt{Cappellari2023,Scholtz2024}), the smooth large-scale variation of the spectrum was modeled as a multiplicative polynomial of 11th order in the fitting run. Other details of the fitting setup are similar to what is discussed in \cite{Wang2024}.  

A SED model was generated from the best-fit parameters and is shown in Figs.~\ref{fig:nirspec_w_models} \& \ref{fig:prospector_full}, along with the best-fit SFH in the right panel of Fig.~\ref{fig:prospector_full}. The best-fit SFH, albeit subject to substantial uncertainties, suggests that the star formation rate started to decline from a few hundred Myrs prior to the epoch of observation.

\begin{figure*}
\centering
\includegraphics[width = 7.2in]{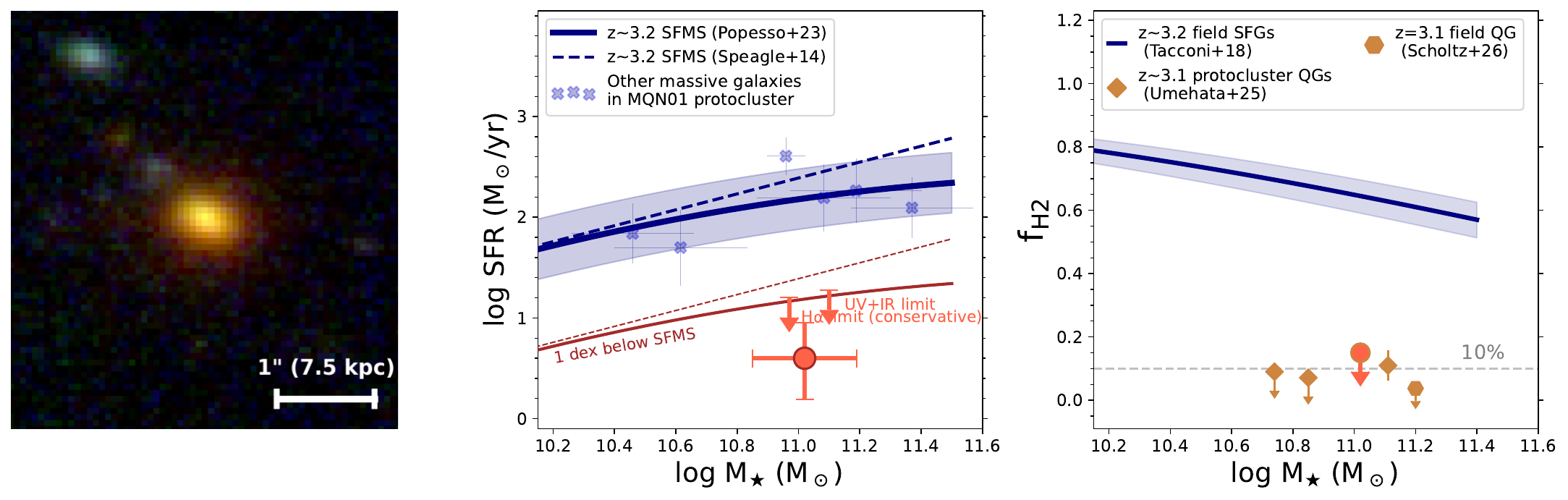}
\caption{Morphology and physical properties of the Red Potato galaxy, which is marked by large red circles in the middle and right panels. \emph{Left:} The galaxy has a red and compact morphology in the JWST+HST color image. \emph{Middle:} It is a massive system with $M_\star \simeq 10^{11}~M_\sun$ and at least one dex below the main sequence at its redshift of $z=3.250$ (blue curves; \citealt{Speagle2014,Popesso2023}), according to the SED fitting and the SFR upper limits from UV+IR and H$\alpha$ (red downward arrows). Its SFR is also substantially lower than those of other massive galaxies discovered in the same protocluster MQN01 (light blue squares; see \citealt{Galbiati2025,Pensabene2025} and main text). \emph{Right:} The galaxy is poor in molecular gas, with a gas fraction $f_\mathrm{H_2}\lesssim 0.15$, substantially smaller than those of the field star-forming galaxies (SFGs, solid curve; \citealt{Tacconi2018}). Similarly low fractions are also found in other massive $z\sim 3$ quiescent galaxies (\citealt{Umehata2025b,Scholtz2024}). The $f_\mathrm{H_2}$ values of the literature points have been recalculated in a way consistent with this work, namely adopting the $r_{31}$ value from \mbox{\cite{Birkin2021}} and $\alpha_\mathrm{CO} = 4\ M_\sun~\mathrm{K^{-1}~km^{-1}~s~pc^{2}}$, for the purpose of direct comparison. All upper limits in the figure are 2-sigma limits.} \label{fig:scaling_relations}
\end{figure*}

The Red Potato galaxy is a massive system with a stellar mass of $1.1^{+0.4}_{-0.4} \times 10^{11}~M_\odot$ and a SFR of $4^{+6}_{-2}~M_\odot$/yr (100-Myr timescale) according to the fitting results, where the quoted errors include both the 1-$\sigma$ statistical uncertainties measured from the fitting posteriors and the systematic errors quantified by \cite{Pacifici2023}. 
As shown in the middle panel of Fig.~\ref{fig:scaling_relations}, the measured stellar mass and SFR values place the galaxy more than 1 dex below the SFMSs reported by the literature \citep{Popesso2023,Speagle2014}. Other massive ($>10^{10}~M_\sun$) galaxies discovered in the same protocluster MQN01  from \cite{Galbiati2025} and \cite{Pensabene2025} are also shown in the panel for comparison, which are all located around the main sequence. These objects include the host galaxy of the AGN-\emph{ID2} shown in Fig.~\ref{fig:rgb_w_lya} at $10^{11.2}\,M_\sun$, the Big Wheel galaxy from \cite{Wang2025} at $10^{11.4}\,M_\sun$, and the QSO companion galaxy from \cite{Pensabene2025} at $10^{11.0}\,M_\sun$.

We also tested whether the SED fitting results are robust against the choice of SFH. Specifically, we switched to a parametric SFH, the delayed exponentially declining form SFR($t$)$\,\propto t\cdot e^{-t/\tau}$, and repeated the fitting with identical setups as above for all other parameters. The inferred mass and SFR values are $1.1^{+0.5}_{-0.4} \times 10^{11}~M_\odot$ and $7^{+7}_{-4}~M_\odot$/yr, respectively, consistent with those inferred from the fitting with a nonparametric SFH if taking into account the uncertainties.

In addition to the SED fitting, the low SFR of the galaxy is also supported by measurements of the Balmer lines and the dust continuum emission. The H$\alpha$ line flux measured from the continuum-subtracted spectrum is $5.6 \pm 0.4 \times 10^{-18}$~erg/s/cm$^2$ in the rest frame. If the line were entirely powered by star formation, we can infer from the flux the SFR (10-Myr timescale) to be $2.8 \pm 0.2$~M$_\sun$/yr, following \cite{Kennicutt2012}. If assuming a \cite{Calzetti2001} dust law, we inferred the H$\alpha$ dust attenuation to be 0.5~mag with a 2-$\sigma$ upper limit of 1.8~mag from the H$\alpha$/H$\beta$ line ratio. Adopting this limit value, the corresponding 2-$\sigma$ upper limit of the total SFR is 16~M$_\sun$/yr. We stress that these are highly conservative upper limits, given the presence of AGN radiation. 
The SFR limit can also be evaluated from ALMA observations of the dust continuum, in which the galaxy is not detected at 1.2~mm.  Including the UV SFR measured from the HST F625W flux, this leads to a 2-$\sigma$ upper limit of 19~M$_\sun$/yr for the total SFR if adopting the average IR SED template of submillimeter galaxies (SMGs) by \cite{Dudzeviciute2020} and the IR-to-SFR conversion relation by \cite{Kennicutt2012}. Both of the SFR upper limits calculated above are consistent with the value from SED fitting and indicated in Fig.~\ref{fig:scaling_relations}.

The galaxy nebular line emission is due to AGN radiation according to the BPT diagram criteria \citep{Baldwin1981,Kewley2001}, demonstrated by the galaxy-integrated measurements in Fig.~\ref{fig:nirspec_zoomins} (right panel). The emission line fluxes were measured from the continuum-subtracted NIRSpec spectrum. We discuss the origin of the AGN-type radiation in Sect.\ \ref{sec:discussion}.

\subsection{Low cold gas fraction traced by CO and Na~D}
\label{subsec:analysis_fgas}

The Red Potato galaxy is not detected in CO(4-3) line emission by deep ALMA observations, which can be used to put an upper limit on its cold molecular gas mass ($M_\mathrm{H_2}$). We calculated this upper limit using the sensitivity of the CO(4-3) line flux measurement (Sect.\ \mbox{\ref{subsec:xray_radio_submm}}) and a CO line ratio $r_{41}=L'$(CO~4-3)$/L'$(CO~1-0)$\,=0.34\pm0.04$ measured by \mbox{\cite{Birkin2021}}. The inferred 2-$\sigma$ upper limit of  $M_\mathrm{H_2}$ is $2.0\,(0.4) \times 10^{10} M_\sun$ if assuming $\alpha_\mathrm{CO} = 4\,(0.8)\ M_\sun~\mathrm{K^{-1}~km^{-1}~s~pc^{2}}$ (see \citealt{Bolatto2013}). Such a limit corresponds to a 2-$\sigma$ limit $f_\mathrm{H_2}=M_\mathrm{H_2}/(M_\star+M_\mathrm{H_2})$ of only 0.15 (0.03) for the molecular gas fraction. We note that such limits are conservative estimates: Substantially lower will be obtained if adopting another $\alpha_\mathrm{CO}$ value widely adopted in the literature \mbox{\citep{Boogaard2020,Pensabene2024}}, namely $0.61\pm0.13$.

These constraints indicate that the galaxy is a molecular gas-poor system: as demonstrated in the right panel of Fig.~\ref{fig:scaling_relations}, its molecular gas fraction is around ten times lower than those of typical field star-forming galaxies at similar masses and redshifts measured by \cite{Tacconi2018}, and yet rather consistent with recent measurements of $z\sim3$ quiescent galaxies \citep{Magdis2021,Umehata2025b,Scholtz2024}. The $f_\mathrm{H_2}$ values of the literature points in the figure have been recalculated in a way consistent with this work, namely adopting the $r_{31}$ value from \mbox{\cite{Birkin2021}} and $\alpha_\mathrm{CO} = 4\ M_\sun~\mathrm{K^{-1}~km^{-1}~s~pc^{2}}$, for the purpose of direct comparison.

The gas-poor nature of the galaxy is also corroborated by the nondetection of the Na\,D line absorption tracing the cold atomic gas ($\lesssim 10^3\,$K; \citealt{Veilleux2005}) of the interstellar medium (ISM). The nondetection is demonstrated in the left panel of Fig.~\ref{fig:nirspec_zoomins} and described in Sect.\ \ref{subsec:analysis_kinematics_outflows}.

\subsection{Stellar kinematics and the nondetection of cold outflows}
\label{subsec:analysis_kinematics_outflows}

The stellar kinematics of the galaxy were modeled by fitting the integrated galaxy spectrum with the {\sc ppxf} code \citep{Cappellari2023}. Stellar population synthesis models from the Flexible Stellar Population Synthesis (FSPS; \citealt{Conroy2009,Conroy2010}) were adopted, and the NIRSpec instrument line spread function was taken into account. Shown as the cyan curve in the lower panels of Fig.~\ref{fig:nirspec_w_models}, the best-fit spectrum (excluding emission line components) reproduces the strong absorption features as observed and is consistent with the model spectrum from {\sc prospector} (orange dashed curve therein).

\begin{figure*}[h]

\centering 
\includegraphics[width = 7.2in]{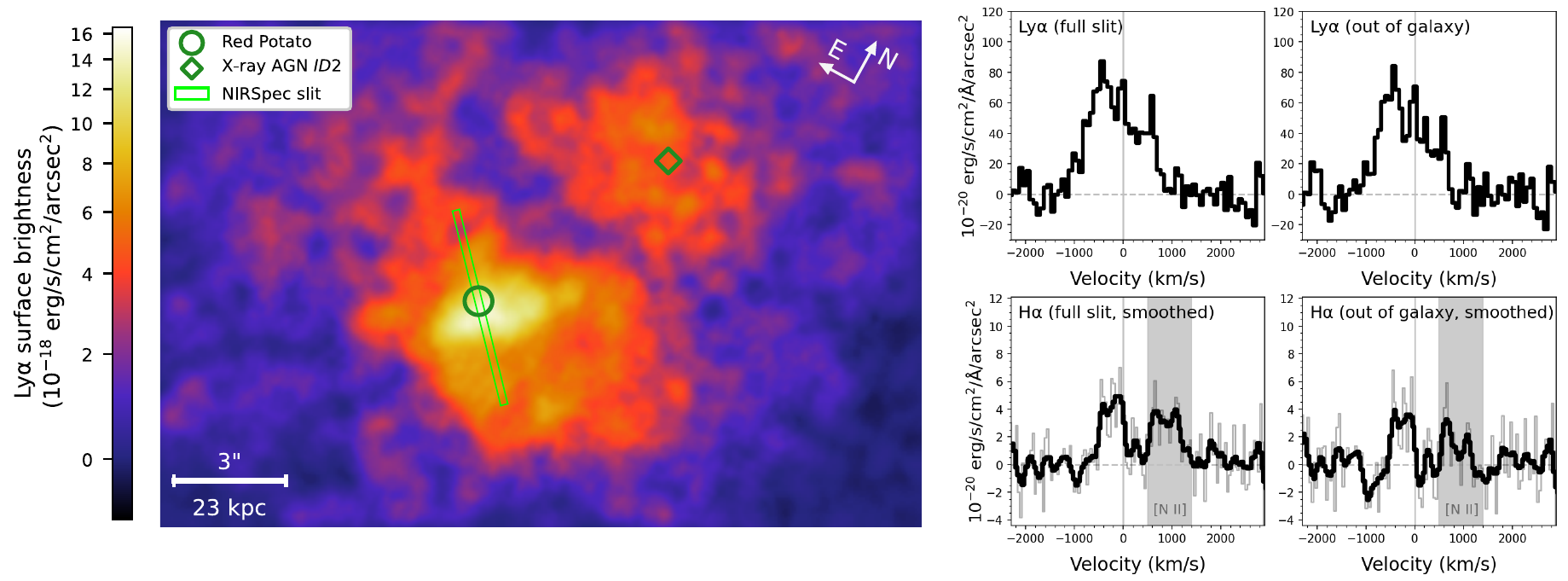}
\caption{Ly$\alpha$ surface brightness map and emission line profiles extracted from the NIRSpec slit, indicating the presence of an extended cool CGM gas reservoir around the Red Potato. \emph{Left panel}: The MUSE Ly$\alpha$ map overlaid with the NIRSpec slit footprint. The Ly$\alpha$ emission is spatially extended with a bright, elongated core reaching a few times $10^{-17}~$erg/s/cm$^2$/arcsec$^2$. \emph{Middle and right panels}: The Ly$\alpha$ (top) and H$\alpha$ (bottom) line profiles extracted from the spatial region covered by the slit. The H$\alpha$ line profile at the native spectral resolution is shown in light gray and the profile after smoothing with a boxcar filter of 4 pixels is shown in black. The gray shade marks the region that could be affected by the [N II]~6583Å line. \emph{Right panels}: Similar to the middle panels but the spectra are extracted from the slit region away from the galaxy, at distances more than 0.35\arcsec~from the galaxy center, which is the slit area not enclosed by the open circle symbol in the left panel. 
The Ly$\alpha$/H$\alpha$ ratio measured from the middle panels is $6.8^{+1.5}_{-1.0}$ and that from the right panels $9.8^{+4.8}_{-2.5}$. These values are both consistent with the recombination emission mechanism considering the possible effect of local Ly$\alpha$ scattering \citep{Langen2023}. Both the Ly$\alpha$ and H$\alpha$ spectral profiles appear relatively broad or present multiple possible components, suggesting complex kinematics and/or a turbulent medium.   
\label{fig:lya_map_fullslitprof}}

\end{figure*}

The stellar velocity dispersion integrated over the whole galaxy and corrected for instrument resolution was measured to be 268$\pm$20 km/s, which is typical of high-redshift quiescent galaxies with stellar masses around $10^{11}\,M_\sun$ \citep{vandeSande2013,Forrest2022,Kriek2024}. The galaxy dynamical mass was estimated to be around $1.1\times10^{11}~M_\odot$ if adopting equation 1 from \cite{Belli2017} and a Sérsic index of 2, which places the Red Potato well within the distribution of other high-redshift massive quiescent galaxies on the dynamical mass--stellar mass diagram presented in \cite{Ito2026}. As for the ionized gas kinematics of the galaxy, the corresponding analysis is deferred to Sect.\ \ref{subsec:analysis_line_spatial_extents2}. 

No molecular and cold atomic gas outflows are detected from the galaxy. The two gas phases are traced by the CO line and Na D doublet lines, respectively. The nondetection of the molecular gas is naturally justified by the nondetection of the CO line (Sect.\ \ref{subsec:analysis_fgas}). Constraints on the neutral outflow are demonstrated in the left panel of Fig.~\ref{fig:nirspec_zoomins}, showing that after removing the best-fit stellar continuum from {\sc prospector} (orange dashed line) the NIRSpec spectrum (black line) does not show any residual absorption of Na D. The same finding is reached if using alternatively the best-fit stellar continuum from {\sc ppxf} (cyan line); masking out the Na D line region or not in the fitting leads to identical results. These analyses suggest that there exists no component of Na D gas absorption or CO gas emission from either the ISM gas or outflowing gas with any velocities or orientations along the line of sight.

The molecular and neutral phases are reported to dominate the mass budget of the AGN- or star formation-driven outflows from typical massive galaxies \citep{Roberts-Borsani2020,Avery2022,Baron2022,Belli2024}. Hence the absence of these mass-loaded outflows from the Red Potato suggests that there is no substantial ejective feedback. In comparison, strong neutral outflows traced by optical/UV absorption lines have been observed in several other $z>3$ massive quiescent galaxies (e.g., \citealt{Man2021,DEugenio2024,Valentino2025,Wu2025,Scholtz2024}; see also \citealt{Belli2024}), likely suggesting that they are at a quenching stage earlier than the Red Potato \citep{Park2024}.

\begin{figure*}
\centering 
\includegraphics[width = 7.5in]{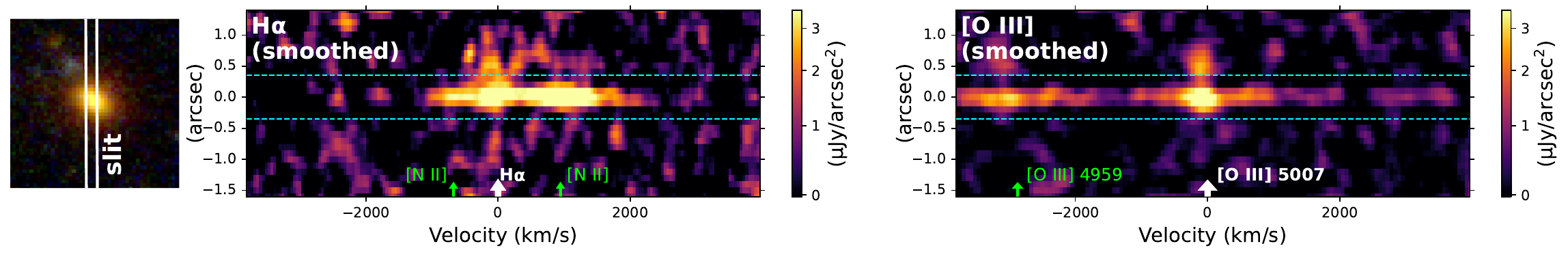}
\caption{Spectra of H$\alpha$ and [O III] 5007\AA\ lines tracing the extended cool ionized gas around the Red Potato. \emph{Left:} The NIRSpec slit footprint from where the 2D spectra were extracted. 
\emph{Middle and Right:} The 2D spectra of H$\alpha$ and [O III]. Both spectra were aligned spatially with the slit image along the Y axis (in unit of arcsec), and the galaxy spatial extent and locations of the nearby emission lines are all marked in each panel. Both emission lines extend beyond the galaxy spatial extent along the slit. 
The spectra were smoothed along the Y direction to a spatial resolution of 0.7\arcsec~ (FWHM), and along the X axis with a boxcar filter of 4 and 3 pixels for H$\alpha$ and [O III], respectively, corresponding to 200\,km/s. The continuum has been removed 
through a row-by-row median filtering. \label{fig:halpha_lya_2dspec}}

\end{figure*}

\begin{figure*}

\centering 
\includegraphics[width = 7.2in]{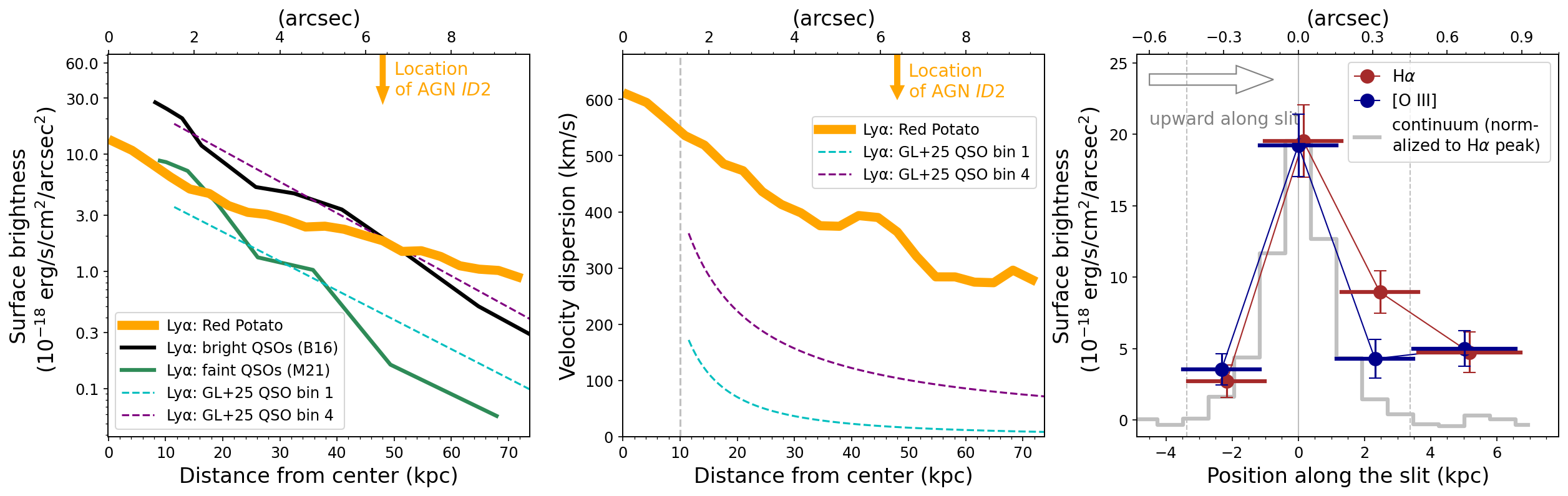}
\caption{Spatial extent and kinematics of Red Potato's CGM measured from emission lines.  
\emph{Left:} Radial Ly$\alpha$ surface brightness profile measured around the Red Potato (orange curve). The Ly$\alpha$ emission is remarkably bright and comparable to those of bright QSOs at $z\sim3$. The black curve represents the median of the Ly$\alpha$ nebulae around bright $z\sim3$ QSOs from \citet{Borisova2016} and the green curve for faint $z\sim3$ QSOs from \citet{Mackenzie2021}. Red Potato's profile is flatter than the QSO profiles at $r\gtrsim30~$kpc, possibly due to the contribution of Ly$\alpha$ emission around the nearby AGN-\emph{ID2}, whose location is marked with the orange arrow. The dashed lines represent the best-fit profiles for two subsamples of $z\sim3$ QSO nebulae from \cite{GonzalezLobos2025} which bracket Red Potato's profile at $r<30~$kpc. The QSO Ly$\alpha$ surface brightness values are all adjusted to Red Potato's redshift. 
\emph{Middle:} The Ly$\alpha$ velocity dispersion profile measured around the Red Potato (orange curve). Measurements for the two subsamples of \cite{GonzalezLobos2025} are shown for comparison (dashed lines). Red Potato's CGM has substantially higher Ly$\alpha$ velocity dispersion than the two subsamples at 20~kpc$\;\lesssim r \lesssim30$~kpc. 
\emph{Right:} Surface brightness of H$\alpha$ and [O III]~5007\AA\ along the NIRSpec slit. The emission lines are more extended than the continuum (gray) along the slit. The continuum is normalized to the H$\alpha$ profile peak. 
\label{fig:radial_profiles}}

\end{figure*}

\subsection{Galaxy morphology}
\label{subsec:analysis_morphology}

In terms of morphology, the galaxy appears as a red and compact spheroid as shown in the left panel of Fig.~\ref{fig:scaling_relations}. The image was created using the F322W2 (3.2~\micron), F150W2 (1.5~\micron), and F814W (0.8~\micron) filters for the R, G, B channels, respectively.
We also analyzed the morphology quantitatively by fitting the NIRCam images with Sérsic profiles \citep{Sersic1963} using the {\sc pysersic} code \citep{Pasha2023}, which is demonstrated in Appendix \ref{sec:appendix_sersic}. The galaxy is best fit with a half-light radius of 0.13 and 0.11 arcsec and a Sérsic index of 1.5 and 2.6 for the F150W2 and F322W2 filters, respectively. 
The galaxy half-light radius in the rest-frame optical (0.5~\micron) can be inferred by interpolating between the F150W2 and F322W2 sizes (\citealt{Wang2025}). The inferred radius is 0.9 kpc, which is consistent with those of the field quiescent galaxies at similar masses and redshifts and substantially smaller than those of star-forming galaxies \citep{Baker2025,Kawinwanichakij2025,Yang2025}. The Sérsic index values (1.5, 2.6) are smaller than the typical value of 4 as found for the local early-type galaxies (e.g., \citealt{Gadotti2009}). Sérsic indices lower than 4 are also reported among other quiescent galaxies discovered at $z>3$, indicating a probable evolutionary connection with disk galaxies at even earlier epochs (e.g., \citealt{Man2021,Carnall2023,Carnall2024,Glazebrook2024,Ito2024,Sato2024,Setton2024,Baker2025,Kawinwanichakij2025}).

\section{Analysis of the CGM and environment} 
\label{sec:measurements_env}

\subsection{Detection of an extended reservoir of cool gas from the Ly$\alpha$, H$\alpha$, and [O III] emission lines}
\label{subsec:analysis_line_spatial_extents}

Extended Ly$\alpha$ emission is found around the Red Potato, indicating the presence of an extended reservoir of cool gas, which is also shown by the Ly$\alpha$ map in Figs.~\ref{fig:rgb_w_lya} \& \ref{fig:lya_map_fullslitprof}. This discovery is corroborated by further measurements of the Ly$\alpha$, H$\alpha$, and [O III] lines. We present these measurements and an analysis of the physical origin of the line emission in the following. 

The Ly$\alpha$ surface brightness map around the Red Potato galaxy is shown in Fig.~\ref{fig:lya_map_fullslitprof}.  The Ly$\alpha$ emission is spatially extended and bright, reaching a surface brightness of a few times $10^{-17}~$erg/s/cm$^2$/arcsec$^2$ in the center. The two upper right panels show the Ly$\alpha$ spectra extracted from a pseudo-slit of the MUSE dataset matched to the NIRSpec slit, and the two bottom right panels show the H$\alpha$ spectra extracted from the NIRSpec slit. 

The Ly$\alpha$/H$\alpha$ line flux ratio measured from the full slit is $6.8^{+1.5}_{-1.0}$. If excluding the galaxy region, namely within 0.35\arcsec~from the galaxy center,  the line ratio measured from outside of the galaxy is $9.8^{+4.8}_{-2.5}$. 
For Ly$\alpha$, the line flux was measured from a velocity window of [-1500, 1500]~km/s considering the broad line profile seen in Fig.~\ref{fig:lya_map_fullslitprof}. For H$\alpha$, the flux was measured from a velocity window of [-500, 500]~km/s, which excludes the region that could be contaminated by the [N II]~6583\AA\ line, as marked in the figure. The H$\alpha$ can be subject to blending with the nearby [N II]~6548\AA\ line, which is three times weaker than the [N II]~6583\AA\ line. To assess the impact of the line blending, we scaled down the emission within the [N II]~6583\AA\ region by a factor of 3, shifted it to the location of the [N II]~6548\AA, and subtracted it from the observed spectrum before measuring the H$\alpha$ line flux. The resulting H$\alpha$ line surface brightness averaged within the slit was measured to be $2.25\pm0.41\times 10^{-18}$~erg/s/cm$^2$/arcsec$^2$, which only decreased by 9\% compared to the original value of  $2.38\pm0.41\times 10^{-18}$~erg/s/cm$^2$/arcsec$^2$, indicating no substantial impact of possible line blending on the H$\alpha$ flux measurement.

The high surface brightness of the extended Ly$\alpha$ and the measured  Ly$\alpha$/H$\alpha$ ratios support a scenario for which the emission lines are produced by recombination radiation of a highly clumpy medium in and around the galaxy (e.g., \citealt{Cantalupo2014,Cantalupo2019, Leibler2018, Langen2023}).
We also present the 2D spectra of H$\alpha$ and [O III] 5007\AA\ and show that they extend beyond the galaxy in Fig.~\ref{fig:halpha_lya_2dspec}. The slit footprint is shown in the left panel and aligned with the 2D spectra along the Y axis. 
Both lines extend beyond the NIRSpec continuum (Sect.\ \ref{subsec:nirspec_muse}), which is marked by the horizontal dashed lines. Specifically, the  H$\alpha$ and [O III] lines reach as far as 1 arcsec or 8 kpc above (i.e., north to) the galaxy center. Tentative H$\alpha$ detection is also found below the galaxy.

\subsection{Detailed analysis of CGM spatial extent and kinematics}
\label{subsec:analysis_line_spatial_extents2}

We present a detailed analysis of the spatial extent of the emission lines and gas kinematics in Fig.~\ref{fig:radial_profiles} and show that the emission lines tracing Red Potato's CGM have broad profiles, corresponding to high velocity dispersion. Relevant implications on the CGM turbulence will be discussed later in Sect.\ \ref{sec:discussion}.

Regarding the spatial extent, the Ly$\alpha$ emission around the Red Potato galaxy is found to reach over 50 kpc in radial distance, which is shown in the left panel of Fig.~\ref{fig:radial_profiles}. The Ly$\alpha$ emission is also remarkably bright, only slightly fainter than the average profile of the Ly$\alpha$ nebulae around bright $z\sim 3$ QSOs measured by \cite{Borisova2016} (black curve; $i_\mathrm{QSO}<19\,$mag) and brighter than those of faint $z\sim 3$ QSOs measured by \cite{Mackenzie2021} (green curve; $i_\mathrm{QSO}\gtrsim20\,$mag). Red Potato's Ly$\alpha$ surface brightness profile is also flatter than the QSO profiles at $r\gtrsim30~$kpc due to the contribution of Ly$\alpha$ emission around the nearby AGN-\emph{ID2}, whose location is marked with the orange arrow in the figure.

The spatial extents measured from H$\alpha$ and [O III] lines along the NIRSpec slit are also shown in  Fig.~\ref{fig:radial_profiles} (right panel). The H$\alpha$ and [O III] flux profiles (red and blue curves) are both more extended than the galaxy continuum (gray curve). The H$\alpha$ appears more extended than the [O III] toward the upward direction along the slit, indicating that the former might be the better tracer of the large-scale CGM for the case of the Red Potato. 

Regarding the gas kinematics, the Ly$\alpha$ velocity dispersion of Red Potato's inner CGM (20~kpc$\;\lesssim r \lesssim30$~kpc) is  400--500 km/s, which is significantly higher than those measured from the $z\sim 3$ QSOs with comparable Ly$\alpha$ surface brightness at the same distances, which are around 50--200 km/s. This is demonstrated in the middle panel of Fig.~\ref{fig:radial_profiles}. The QSO nebula profiles for comparison are from two subsamples in a study by \cite{GonzalezLobos2025} which bracket the Ly$\alpha$ surface brightness profile of Red Potato's inner CGM, as shown in the left panel. Specifically, these two subsamples are the first and fourth bins listed in the top panel of tab.~3 of the paper. The Ly$\alpha$ velocity dispersion values are all measured from the second moment of the line spectral profile, which is the common practice among the literature (e.g., \citealt{Borisova2016}).
We note that multiple spatial components oriented along our line of sight could also result in an apparently large velocity dispersion. However, we have no indications from the galaxy spatial and redshift distribution that this could be the case, although deeper JWST data would be needed to fully exclude this hypothesis.

We also analyzed the gas kinematics by extracting 1D emission line spectra from regions in and around the galaxy, which are presented in Appendix \ref{sec:appendix_1dspec}.  Analysis of the H$\alpha$ within the galaxy suggested a gas velocity dispersion of around 200 km/s, which is two to three times the typical values measured from SFGs at similar masses and redshifts \citep{Simons2017,Danhaive2025,Wisnioski2025}, with no significant velocity gradient. This indicates that the Red Potato is a dispersion-dominated system. In addition, the H$\alpha$ line profile measured from a region around the galaxy appears broad and complex, which we discuss further in Sect.\ \ref{sec:discussion}.

\begin{figure*}
\centering 
\includegraphics[width = 7.0in]{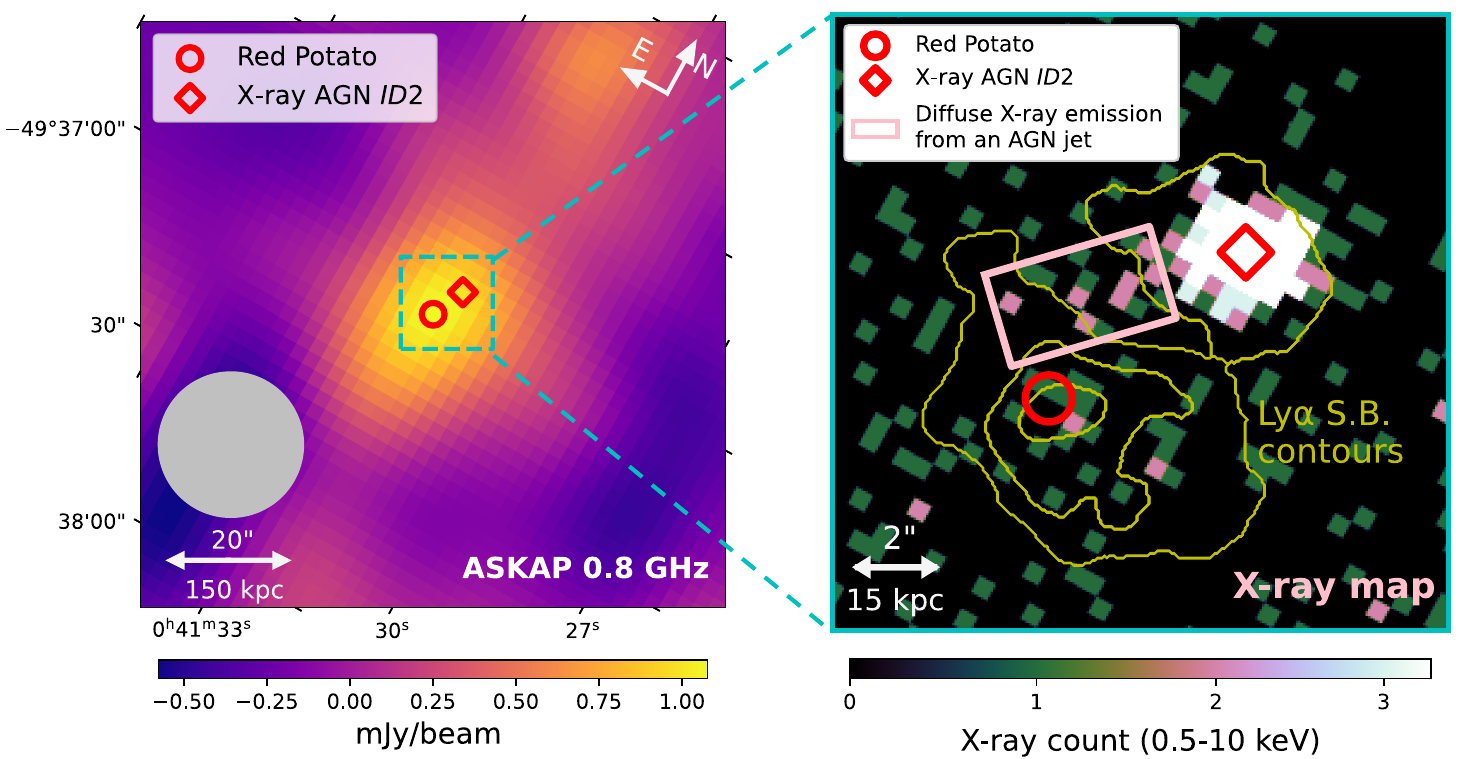}
\caption{Analysis of radio and X-ray emission near the Red Potato galaxy. In the left panel, radio emission at 0.8 GHz is detected at significant levels but not spatially resolved, given the ASKAP beam size of around 20 arcsec (gray circle at the lower left corner). The background image in the right panel is a zoomed-in view of the \emph{Chandra} X-ray count map. Diffuse X-ray emission around the AGN neighbor \emph{ID2} is detected and marked by the pink box, indicating the presence of an X-ray jet which most likely emanates from AGN-\emph{ID2}. The low photon count does not allow a detailed morphological and spectral analysis but the map clearly shows an X-ray photon excess along the direction roughly toward the Red Potato.
The Ly$\alpha$ contours from Fig.~\ref{fig:rgb_w_lya} are also repeated in the right panel for reference. See Sect.\ \ref{subsec:analysis_xray_radio} and \ref{subsec:discussion_quenching_galaxy}.}

\label{fig:nirspec_xray_radio_maps}

\end{figure*}

\subsection{The nearby X-ray AGN and evidence for a jet}
\label{subsec:analysis_xray_radio}

As shown in Figs.~\ref{fig:rgb_w_lya} \& \ref{fig:nirspec_xray_radio_maps}, the Red Potato galaxy is found close to another galaxy which is at a similar redshift of $z=3.251$ (versus $z=3.250$ for the Red Potato) and hosts a luminous X-ray AGN \citep{Pensabene2024,Galbiati2025,Travascio2025}. The X-ray source, identified and named as \emph{ID2} in \cite{Travascio2025}, is located at a projected distance of 6.4 arcsec or 48 kpc and has a luminosity of $L_\mathrm{2-10~keV}=10^{45}$ erg/s, making it the brightest X-ray source of the MQN01 structure after the central QSO of MQN01. 
As also shown in the left panel of Fig.~\ref{fig:nirspec_xray_radio_maps}, the Red Potato galaxy position, as well as the AGN-\emph{ID2}, is within the 20"-wide ASKAP beam of a radio source detected at 0.8 GHz, indicating a certain form of radio AGN feedback in action. The source has a measured flux of $1.56 \pm 0.47$ mJy, the brightest of the large-scale structure.  There is no detection at 1.4 GHz, corresponding to a 2-$\sigma$ flux upper limit of 0.36 mJy (\citealt{Duchesne2024}). From the 0.8 GHz and 1.4 GHz measurements, a radio spectral index $\alpha$, defined as $S\propto \nu^{-\alpha}$ where $S$ is the flux density and $\nu$ is the frequency, can be determined. The inferred $\alpha$ is above 2.8 at 2-$\sigma$ significance. Such a steep spectral index can be produced by an aged radio jet \citep{Myers1985,Alexander1987,Brienza2021}.

The right panel of Fig.~\ref{fig:nirspec_xray_radio_maps} shows the X-ray count map, which is zoomed onto a smaller sky region. Intriguingly, the X-ray emission is found extended over about 40 kpc, which is marked by the pink box in the figure, and the extended emission is approximately aligned in the direction from the AGN-\emph{ID2} to the Red Potato. The minimal distance from the box region to AGN-\emph{ID2} is 2 arcsec, which is the 98\% photon-enclosing radius of the Chandra PSF \mbox{\citep{Travascio2025b}}, so contamination by the emission from \emph{ID2} itself is negligible. We determined the significance of this extended emission by measuring the distribution of photon counts from 5000 background apertures with the same area as the box; the apertures are randomly placed in a region within 1.6 arcmin from the Chandra aim point \mbox{\citep{Travascio2025}} to ensure the PSF vary less than 10\% across the region \mbox{\citep{Tozzi2022}}. The photon distribution has a median of 6.7 and a standard deviation of 2.5. The extended structure has 22 photons measured, thus corresponding to a significance of about 6-$\sigma$. Such emission is indicative of an AGN jet, which most likely emanates from the AGN-\emph{ID2} since the extended emission appears connected to the AGN-\emph{ID2} itself on the X-ray image. This jet is most likely the cause of the radio emission seen in the left panel, although this remains to be examined by high-resolution radio observations in the future.
Further discussions regarding the nature of this jet and its impacts to the Red Potato galaxy are presented in Sect.\ \ref{subsec:discussion_quenching_galaxy}.

\section{Discussion}
\label{sec:discussion}

\subsection{How to maintain the quiescent state of a galaxy at the center of a large cool gas reservoir at $z>3$}

\label{subsec:discussion_quenching_galaxy}

The most intriguing result of this work is the discovery of a passive galaxy, with a SFR at least one dex below the SFMS, right at the center of a large cool gas reservoir traced by Ly$\alpha$. Furthermore, no molecular gas is detected inside the galaxy, corresponding to a molecular gas fraction limit of $f_\mathrm{H2}<0.15$. 
In the context of the cosmological picture described at the beginning of Sect.\ \mbox{\ref{sec:intro}}, these two results suggest that either gas accreting from the intergalactic medium (IGM) cannot efficiently reach the galaxy over timescales longer than a few hundred Myrs or that some process is reducing the ability of the gas to efficiently accumulate within the galaxy (thus removing one of the necessary conditions for possibly further condensation and efficient star formation).

Important clues to disentangling the two possibilities mentioned above come from a detailed analysis of the CGM properties. 
First, the Ly$\alpha$ SB profiles and the overall morphology of the nebula seem to suggest that cool CGM gas is present even in the innermost halo region, with a Ly$\alpha$ peak just a few kpc away from the photometric center of Red Potato (Fig.~\ref{fig:lya_map_fullslitprof}). Unless the detection of the Ly$\alpha$-emitting gas is due to a chance projection effect, the presence of this gas close to the galaxy appears inconsistent with a scenario in which cool gas around the galaxy is reduced due to reasons such as an ongoing transition from the so-called ``cold mode" to ``hot mode" as predicted by some theoretical models (\citealt{Dekel2006,Dekel2009}).
Second, the spectral analysis of the Red Potato's CGM emission both in Ly$\alpha$ and H$\alpha$ shows broader line profiles than those previously measured for CGM emission around AGN. For instance, the Ly$\alpha$ velocity dispersion in a circular annulus between 20 and 30 pkpc from the Red Potato has values between 400 and 500 km/s. This is significantly higher than the velocity dispersion measured for QSO nebulae with similar surface brightness values which range from 50 to 200 km/s within the same circular annuli, as measured by, for example, \cite{GonzalezLobos2025}, and shown in the middle panel of Fig.~\ref{fig:radial_profiles}. The broadness of the Ly$\alpha$ line could be affected by radiative transfer, the impacts of which can only be fully characterized by a statistical sample of H$\alpha$ spectral profiles; Unfortunately, such a sample is not currently available at z$>$3.
Nevertheless, the H$\alpha$ spectral profile in the CGM of the Red Potato shows complex and broad kinematics (Figs.~\ref{fig:lya_map_fullslitprof} \& \ref{fig:halpha_lya_1dspec}), suggesting that the large velocity dispersion measured from the Ly$\alpha$ spectra cannot be solely due to radiative transfer effects but is also caused by turbulent gas kinematics. 

If sustained over a sufficiently long time, this  turbulence in the CGM surrounding the Red Potato could help prevent the gas from efficiently accumulating within the galaxy, thus removing one of the necessary conditions for possibly further condensation and efficient star formation. We caution that such a scenario is intended for the specific case presented in this work, and that further studies are needed to examine whether it could be applicable to other high-$z$ quiescent galaxies.

After justifying that the galaxy quiescence is more likely caused by the CGM turbulence preventing the gas from efficiently accumulating onto the galaxy (namely, the second scenario described above), we further discuss what could be causing and maintaining this increased level of CGM turbulence.  
One possibility could be linked to the previous history of the Red Potato itself, including ejective feedback episodes related to starbursts or AGN activity within the Red Potato which could have depleted its molecular gas reservoir and, at the same time, produced a turbulent CGM. Currently, there is no clear evidence of the presence of an active AGN in the Red Potato from Chandra or ALMA observations. At the same time, there is no clear evidence of strong outflows from the Red Potato, either in the form of broad or blueshifted components in [OIII] and CO emission lines or in the form of blueshifted absorption in the Na~D lines. As such, although ejective feedback episodes could have initiated the quenching, there is currently no clear evidence that they can maintain the continuous suppression of the Red Potato's star formation with the presence of the cool-gas-rich CGM, over the observed timescales of a few hundred Myrs (Fig.~\ref{fig:prospector_full}).

Instead, a more plausible culprit of the turbulence is the AGN jet. Our deep \emph{Chandra} observations clearly show that there is a currently active and bright AGN in the vicinity of the Red Potato and present substantial evidence of a jet which emanates from this AGN in the direction of the Red Potato. The presence of a jet is also supported by the ASKAP radio observations which, however, currently lack the resolution to study its detailed morphology.  
The presence of a jet and its association with AGN-\emph{ID2} thus suggest that this external cause  could be a probable agent currently maintaining an increased turbulence level in the Red Potato CGM.
Indeed, radio jets are known to be capable of disturbing and dynamically heating gas surrounding galaxies, at least in the local universe, and thus significantly elevate the CGM turbulence (\citealt{Fabian2012}, \citealt{Krause2023} and references therein), leading to a possibly reduced efficiency for the gas to efficiently accumulate within the galaxy.

To our knowledge, AGN jet lifetimes have not been constrained by observational studies at this redshift. 
Studies at $z\sim0$ suggest that AGN jets can emerge from consecutive episodes and remain detectable on the halo scale for at least 200 Myrs \citep{Brienza2025}, which might be comparable to the relevant timescale implied for observing the Red Potato in a quiescent state. We also note that, in order to fully establish this scenario as a viable option, it is required that the AGN-\emph{ID2} remain not too faraway from the Red Potato over a sufficiently long timescale despite the possible relative motions between these two objects, which we defer to investigate in a future work. 
We also note that the host galaxy of AGN-\emph{ID2} is a SFG on the main sequence at this redshift (see \citealt{Galbiati2025}); thus apparently the radio jet is not currently affecting its SFR. This is often seen at high-redshift, where the brightest radio-galaxies are also among the systems with the highest SFR (e.g., the SpiderWeb galaxy; \citealt{Pentericci1997,Pentericci2000,Miley2006}). This is consistent with the turbulent CGM scenario discussed above, in the plausible case that the AGN-\emph{ID2} jet has an opening angle that is too small to significantly affect the inner CGM of its host galaxy but widens on larger spatial scales where it encounters the CGM of Red Potato. Such a configuration of radio jets is often seen in local sources (e.g., \citealt{Brienza2021,Brienza2025}). 

The proximity of AGN-\emph{ID2} to the Red Potato could also have additional implications. For instance, a recent encounter between these two galaxies could have gravitationally perturbed or displaced the ISM of the Red Potato, increasing the velocity dispersion of the gas and thus reducing its star formation efficiency.

\subsubsection{Comparing with literature studies}

The Red Potato adds to the small yet growing collection of $z\gtrsim 3$ quiescent galaxies discovered in overdense, gas-rich environments \citep{Kalita2021,Kubo2021,deGraaff2025,Umehata2025a,Perez-Martinez2025,Guo2025}. Preventive jet-mode AGN feedback has been commonly proposed to explain their existence. 
Two of these studies investigate the impacts of such preventive feedback by analyzing the CGM properties in detail. \cite{PerezGonzalez2025} discover H$\alpha$-emitting CGM with high velocity dispersion values of 200--300~km/s around a quiescent galaxy, which agrees with our measured values (Sect.\ \ref{subsec:analysis_line_spatial_extents}) and indicates the presence of turbulent cool CGM. On the other hand, \cite{Guo2025} discover a large Ly$\alpha$-emitting CGM which is centered at a quiescent galaxy and has a steep radial velocity gradient according to the Ly$\alpha$ line profiles. 
They interpret the Ly$\alpha$ velocity gradient as a potential sign of active gas accretion, barring the impacts of resonance scattering on the Ly$\alpha$ profiles (e.g., \citealt{Verhamme2006}), and further argue that the quenching is most likely due to the impacts of AGN feedback on the galaxy interior rather than the halt of CGM gas accretion. This appears at odds with the findings of this work, reporting no significant gas velocity gradient or signs of gas accretion in the H$\alpha$ kinematics. The cause of this potential discrepancy remains to be investigated with better sample statistics and a more comprehensive analysis of the CGM kinematics using nonresonant emission lines such as H$\alpha$ in the future. 

\subsection{The diverse evolution pathways of massive galaxies in high-redshift overdense environments}
\label{subsec:discussion_formation_pathways}

The findings of this work, along with those from recent literature studies, indicate a broad range of star formation and morphological properties among massive galaxies in high-redshift protoclusters or cosmic web nodes. For example, the target field of this work hosts a high concentration of massive SFGs \citep{Pensabene2024,Galbiati2025}, including the giant disk Big Wheel \citep{Wang2025}, and also massive compact quiescent galaxies such as the Red Potato (see also Galbiati et al.~in prep.). Although located in the same overdense environment and with similar stellar masses, the physical sizes and SFRs of these two types of galaxies differ by as much as one order of magnitude. Similar diversities are reported by studies of other overdense fields at $z\gtrsim 3$ \citep{Kalita2021,Kubo2021,Venkateshwaran2024,deGraaff2025,Umehata2025b,PerezGonzalez2025,Guo2025}. 
These findings suggest that massive galaxies in high-redshift overdense environments have diverse physical properties, and yet it remains elusive why such diversity occurs. 
Follow-up studies are needed to fully resolve this puzzle by investigating statistical samples of massive galaxies in such environments (Galbiati et al.~in prep.).

\section{Summary and implications} \label{sec:summary}

In this work, we present the serendipitous discovery of a massive red galaxy ($M_\star \simeq 10^{11} M_\sun$) located in a gas-rich cosmic web node at $z=3.250$. Surprisingly, the galaxy, which we call the ``Red Potato,'' is quiescent with a SFR at least 1 dex below the SFMS (Sect.\ \ref{sec:measurements}), although it is at the center of an 80-kpc reservoir of cool CGM (10$^4$--10$^5$\,K) traced by the extended Ly$\alpha$ emission (Sect.\ \ref{sec:measurements_env}). 

In terms of gas inside the galaxy, the Red Potato is poor in molecular and neutral gas according to nondetections of the CO(4--3) and Na~D lines, indicating a molecular gas fraction smaller than 10\% (Sect.\ \ref{subsec:analysis_fgas}). Neither does it have detectable gas outflows (Sect.\ \ref{subsec:analysis_kinematics_outflows}). 
Rest-frame optical emission lines are detected from the galaxy, which are consistent with being driven by AGN illumination rather than star formation  (Sect.\ \ref{subsec:analysis_sedfitting_sfr_agn}). The galaxy appears to be a dispersion-dominated system according to the kinematics of ionized gas traced by H$\alpha$.

In terms of CGM properties, the Red Potato is located at the center of a large reservoir of gas traced by a Ly$\alpha$ halo which is around 80 kpc in diameter at a sensitivity limit of $3\times 10^{-18}~$erg/s/cm$^2$/arcsec$^2$ (Sect.\ \ref{subsec:analysis_line_spatial_extents}). In addition, the H$\alpha$ and [O III] lines are also more spatially extended than the galaxy stellar continuum.  
The Red Potato CGM shows high velocity dispersion values according to the observed Ly$\alpha$ and H$\alpha$ line profiles, which indicate elevated levels of gas turbulence in the CGM compared to other Ly$\alpha$ nebulae at similar redshifts and Ly$\alpha$ surface brightness values (Sect.\ \ref{subsec:analysis_line_spatial_extents2}).

Intriguingly, deep X-ray observations suggest the presence of an extended X-ray jet which most likely emanates from a luminous X-ray AGN neighbor, indicating a certain form of jet-mode feedback acting on the Red Potato CGM. This is corroborated by the detection of bright radio emission in the low-resolution ASKAP 0.8 GHz map (Sect.\ \ref{subsec:analysis_xray_radio}). 

To explain how the Red Potato is maintained at its observed low-SFR state, we propose a scenario in which the AGN jet may have led to increased CGM turbulence around the Red Potato, which keeps this cool gas reservoir in a dynamically hot, turbulent state and thus reduces its ability to efficiently accumulate within the galaxy, removing one of the necessary conditions for possibly further condensation and efficient star formation (Sect.\ \ref{sec:discussion}).  Additional or alternative energy injection mechanisms could involve the gravitational interaction between the Red Potato and AGN-\emph{ID2}. In all scenarios, the nearby AGN(s) or the large overdensity of AGNs associated with the Red Potato's environment may also be illuminating the Red Potato's cool CGM component, making it visible through fluorescent line emission. The scenarios described above are supported by the deep multiwavelength observations unique to this field (\citealt{Pensabene2024,Travascio2025,Galbiati2025,Wang2025}).

Our study demonstrates that the star formation rates of high-redshift galaxies could be
substantially reduced and maintained at a low level even within gas-rich and overdense environments in particular situations, such as for galaxies in the vicinity of AGNs with active jets.
We emphasize that, in order to fully resolve the puzzle of early-epoch galaxy quenching, future studies need to target not only the galaxies themselves but also their CGM and environment. This work is one of the few studies to date that have probed the CGM of quiescent galaxies at $z>3$. Future studies utilizing a larger galaxy sample for this purpose would be essential to reveal the quenching pathways of early-epoch galaxies from a statistical perspective.

\begin{acknowledgements} 
WW would like to thank M.~Brienza, C.~D'Eugenio, L.~di Mascolo, Min Du, F.~Fiore, L.~Kimmig, R.~Lico, P.~Oesch, P.~Pérez-González, R.-S.~Remus, Dongdong Shi, I.~Smail, N.~Sulzenauer, and Tao Wang for insightful discussions related to this work.
This project was supported by the European Research Council (ERC) Consolidator Grant 864361 (CosmicWeb). AP acknowledges support from the Independent Research Fund Denmark (DFF) under grant 3120-00043B. This work is based in part on observations made with the NASA/ESA/CSA James Webb Space Telescope. The data were obtained from the Mikulski Archive for Space Telescopes at the Space Telescope Science Institute, which is operated by the Association of Universities for Research in Astronomy, Inc., under NASA contract NAS 5-03127 for JWST. These observations are associated with program \#1835.
Support for program \#1835 was provided by NASA through a grant from the Space Telescope Science Institute, which is operated by the Association of Universities for Research in Astronomy, Inc., under NASA contract NAS 5-03127. This research is based on observations made with the NASA/ESA Hubble Space Telescope obtained from the Space Telescope Science Institute, which is operated by the Association of Universities for Research in Astronomy, Inc., under NASA contract NAS 5–26555. These observations are associated with program 17065.
ALMA is a partnership of ESO (representing its member states), NSF (USA) and NINS (Japan), together with NRC (Canada), MOST and ASIAA (Taiwan), and KASI (Republic of Korea), in cooperation with the Republic of Chile. The Joint ALMA Observatory is operated by ESO, AUI/NRAO and NAOJ. The scientific results reported in this article are based in part on observations made by the \emph{Chandra} X-ray Observatory. This work is also based on observations collected at the European Southern Observatory under ESO program (110.23ZX).
\end{acknowledgements}

\bibliography{references}{}
\bibliographystyle{aa}

\begin{appendix}

\onecolumn 
\nolinenumbers

\section{Flux calibration of the NIRSpec spectrum}
\label{sec:appendix_slitloss}

\begin{figure*}[h]
\sidecaption
\includegraphics[width = 4.75in]{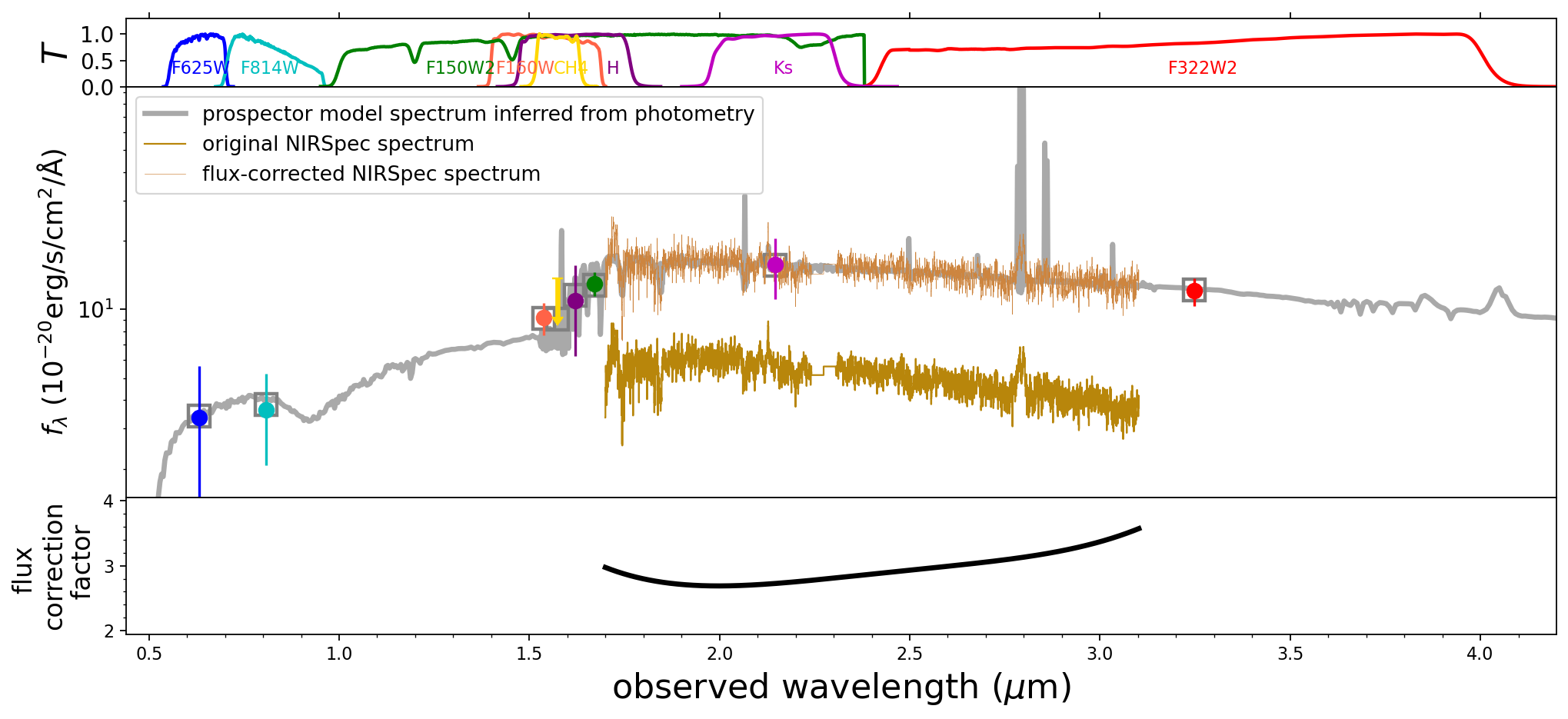}
\caption{Correction of the slit flux loss to the NIRSpec spectrum. The observed spectrum (thick goldenrod curve) has an overall lower flux level compared to the total photometric fluxes (filled circles) because the slit does not cover the entire galaxy. To correct for this, a model spectrum (gray curve) was obtained from an SED fitting of only the multiband photometric fluxes. The ratio between the model spectrum and observed spectrum excluding the emission line regions was fit as a fifth-order polynomial of the wavelength, leading to a wavelength-dependent factor presented in the lower panel. This factor was applied to the observed NIRSpec spectrum before it was used for the SED fitting in Sect.\ \ref{subsec:analysis_sedfitting_sfr_agn}.  \label{fig:prospector_fluxcal} }
\end{figure*}

The NIRSpec slit has a width of 0.2\arcsec, significantly smaller than the spatial extent of the Red Potato galaxy. The loss of flux out of the slit leads to an overall lower flux level of the NIRSpec spectrum, compared to the total fluxes measured from the photometric filters covering the same wavelength range. To address this, we first conducted a {\sc prospector} run using only the multiband photometric data, with other setups identical to those described in Sect.\ \ref{subsec:analysis_sedfitting_sfr_agn}, and obtained a best-fit model spectrum from the fitting, which is shown as the gray curve in the middle panel of Fig.~\ref{fig:prospector_fluxcal}. Then, this model spectrum was compared with the observed NIRSpec spectrum (thick goldenrod curve) to obtain a wavelength-dependent flux correction factor: the ratio between the former and latter was fit as a fifth-order polynomial of the wavelength, leading to a correction factor presented in the lower panel of the figure. An outlier clipping step was implemented in the process to avoid the impact of the misfit emission lines, which were unconstrained in the photometry-only SED fitting. The resulting factor, overall increasing with the wavelength, was multiplied onto the NIRSpec spectrum to account for the flux loss, before it was used for the SED fitting in Sect.\ \ref{subsec:analysis_sedfitting_sfr_agn}.

\section{NIRCam image Sérsic fitting results}
\label{sec:appendix_sersic}

\begin{figure*}

\centering

\begin{minipage}[t]{6.6in} 
    \centering 
    \includegraphics[width = 6.5in]{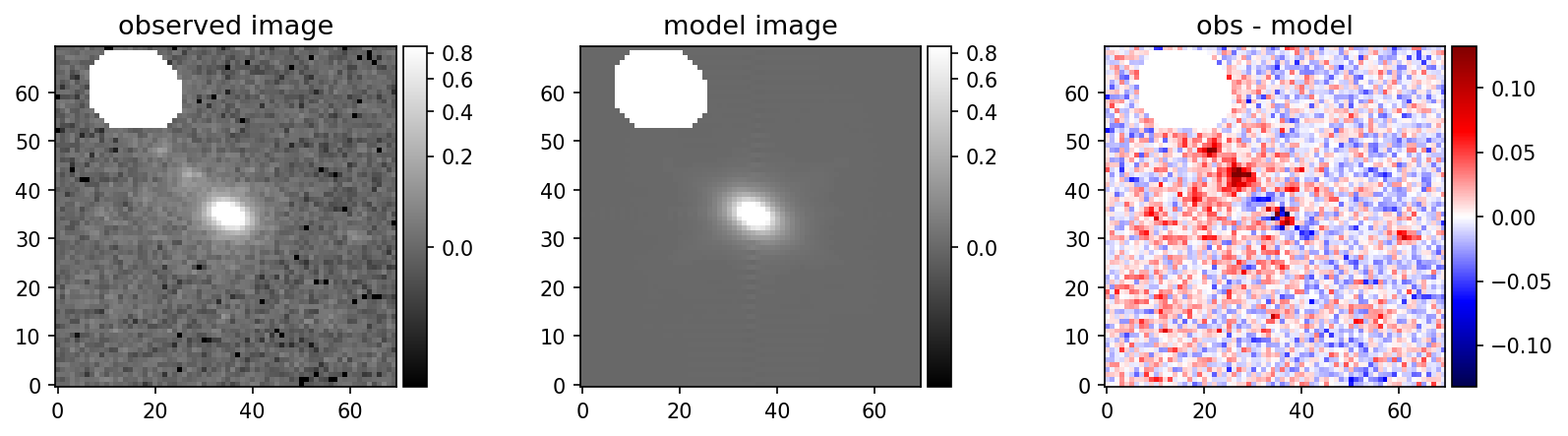}
\end{minipage} 
\hspace{2.5cm} 
\begin{minipage}[t]{6.6in} 
    \centering 
    \includegraphics[width = 6.4in]{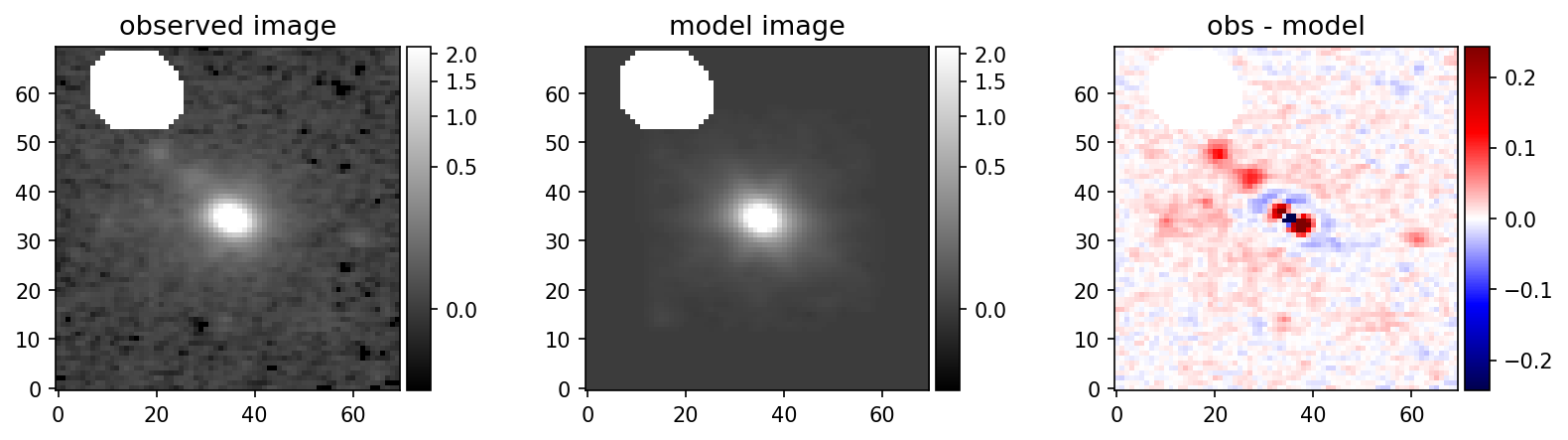} 
\end{minipage} 

\caption{Galaxy images (\emph{left}) and the best-fit Sérsic models (\emph{middle}), for the F150W2 (rest-frame 0.4~\micron; top row) and F322W2 filters (rest-frame 0.8~\micron; bottom row). The Red Potato galaxy is compact in both filters, with best-fit half-light radii of 0.13 and 0.11 arcsec, respectively, and Sérsic indices below 3. The residual images of the fitting are shown in the right panels. \label{fig:pysersic_nircam} }

\end{figure*}

The galaxy images in the two NIRCam filters, F150W2 (rest-frame 0.4~\micron) and F322W2 (rest-frame 0.8~\micron), were fit with 2D Sérsic profiles with pysersic \citep{Pasha2023}. The images and best-fit models are presented in the left and middle panels of Fig.~\ref{fig:pysersic_nircam}. The NIRCam PSF was taken into account in the fitting, and a mask was placed near the upper left corner of the image to mask out a separate object projected on the sky. Other details of the fitting setup can be found in \cite{Wang2025}. According to the fitting results, the galaxy is compact in both filters, with half-light radii of 0.13 and 0.11 arcsec and Sérsic indices of 1.5 and 2.6 for F150W2 and F322W2, respectively. 

The residual images are also shown in the right panels, where small objects are found near the Red Potato galaxy. However, without the coverage from the JWST spectroscopic observations, the redshifts of these neighbors cannot be determined. At the galaxy center, no point source is found from the residual map of either filter, indicating that the central AGN of the galaxy (if there exists one) does not have any noticeable continuum emission in the rest-frame UV/optical.

\section{1D emission line spectra extracted from regions in and around the galaxy}
\label{sec:appendix_1dspec}

\begin{figure*}
\centering 
\includegraphics[width = 6.3in]{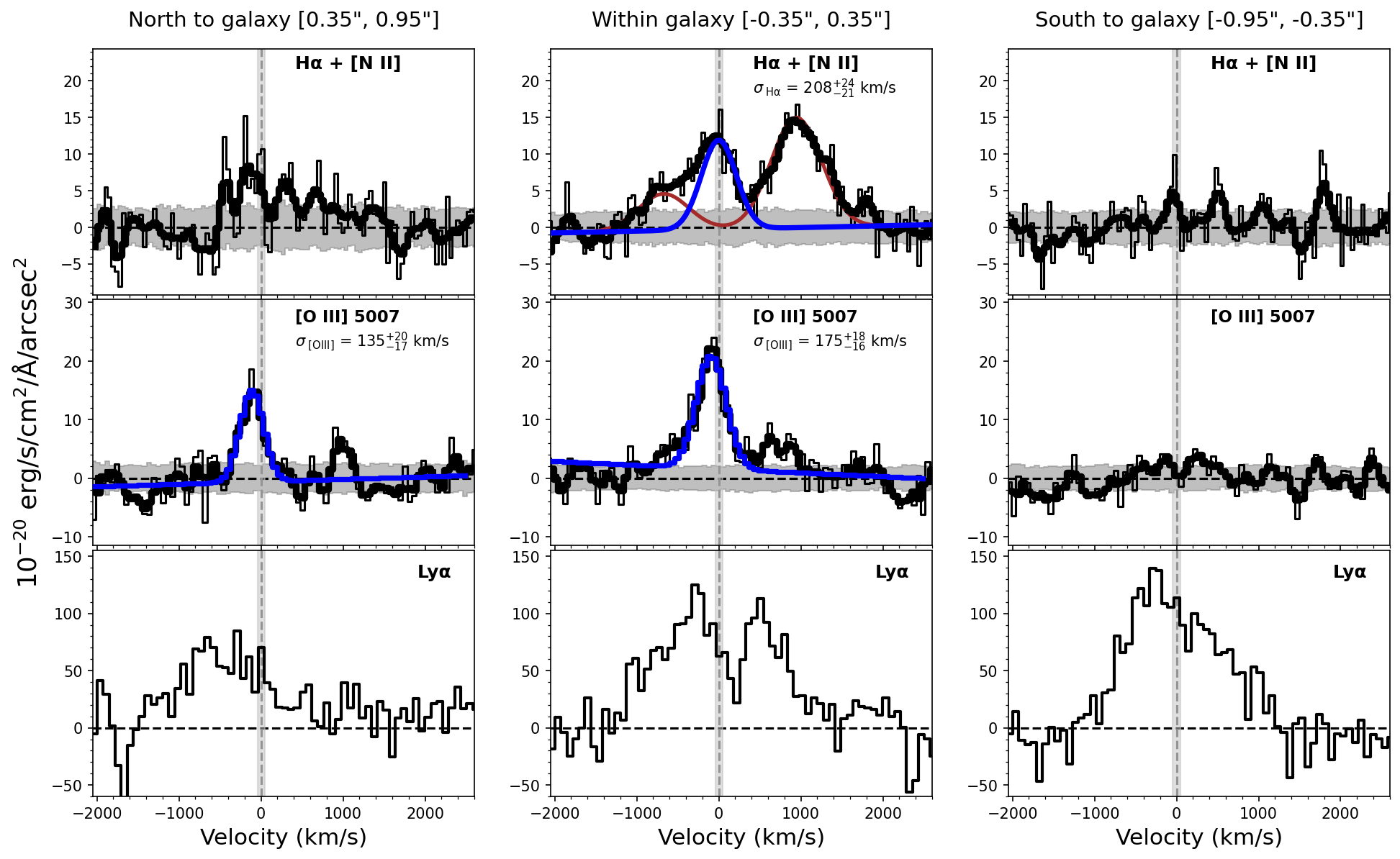}
\caption{Extracted 1D spectra of the H$\alpha$ (top row) and [O III] 5007\AA\ (middle row) from regions in and around the Red Potato along the NIRSpec slit. The Ly$\alpha$ spectra extracted from a wider area around the NIRSpec slit are also presented in the bottom panel for a qualitative comparison (see text for details). From left to right, the spectra are extracted from above (north to), within, and below (south to) the galaxy (c.f.~Fig.~\ref{fig:halpha_lya_2dspec}), respectively. The thin gray lines represent the profiles at the native resolution and the thick black lines represent the smoothed profiles. The H$\alpha$ and [O III] lines are detected in the regions above and within the galaxy. 
Tentative detection of the H$\alpha$ is also found south to the galaxy. The blue curves represent Gaussian profile fitting models of the H$\alpha$ and [O III] lines, when possible. The brown curve represents the fit to the [N II] lines. The velocities are all with reference to the galaxy systemic redshift, the 2-$\sigma$ uncertainty range of which is indicated by the vertical shades. The standard errors of the NIRSpec spectra are indicated as the light gray shades in the top and middle panels. Especially in the northern region, which is the area closer to the X-ray jet detection (see Fig.~\ref{fig:nirspec_xray_radio_maps}), the H$\alpha$ spectrum appears kinematically complex as shown in the top left panel.
\label{fig:halpha_lya_1dspec}}
\end{figure*}

We investigated the gas kinematics further using the 1D spectra of the H$\alpha$ and [O III] lines in Fig.~\ref{fig:halpha_lya_1dspec}.  The spectra were extracted from spatial regions inside the galaxy (middle panels) and around the galaxy (left and right panels). H$\alpha$ and [O III] lines are detected in the regions above and within the galaxy (left and middle panels), and the results of Gaussian fitting to the line profiles are also shown. A tentative detection of H$\alpha$ is also found in the region below the galaxy.  

In the region above the galaxy, both the H$\alpha$ and [O III] lines are detected with a velocity dispersion ($\sigma$; corrected for instrument line spread function) above 100 km/s. In addition, the H$\alpha$ line profile measured from this region appears broad and complex (upper right panel). The gas velocity dispersion measured from the H$\alpha$ line within the Red Potato galaxy is around 200 km/s, which is two to three times the typical values measured from SFGs at similar masses and redshifts \citep{Simons2017,Danhaive2025,Wisnioski2025}, and no significant velocity gradient is found across the galaxy. This indicates that the Red Potato appears to be a dispersion-dominated system, whereas some other high-redshift quiescent galaxies are found to be rotation-dominated (e.g., \citealt{DEugenio2024}). However, more comprehensive spectroscopic observations (e.g., with integral field spectrographs) are needed to make a more conclusive assessment of the dynamical state of ionized gas within the galaxy.

The Ly$\alpha$ line profiles are also shown in the bottom panels of Fig.~\ref{fig:halpha_lya_1dspec} for reference. They are extracted from spatial regions four times wider than the NIRSpec slit area to ensure sufficient signal-to-noise. However, we note that a quantitative comparison between the Ly$\alpha$ and H$\alpha$ lines cannot be performed within these apertures, which are still not large enough to mitigate the impacts of seeing and possible local resonant scattering on the Ly$\alpha$.

\section{ASKAP radio maps}
\label{sec:appendix_askap}

\begin{figure*}
\sidecaption
\includegraphics[width = 5.0in]{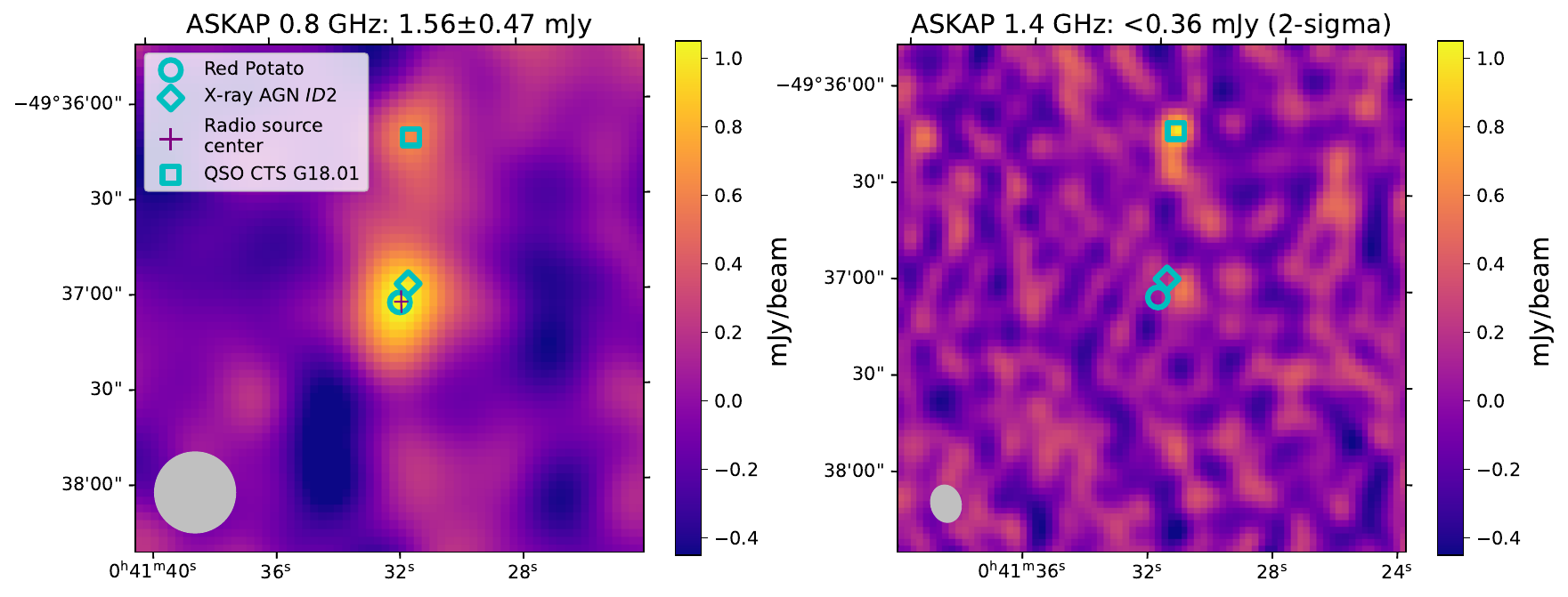}
\caption{ASKAP radio continuum maps around the Red Potato galaxy. \emph{Left:} The galaxy has a radio counterpart in 0.8 GHz, suggesting a certain form of jet-mode feedback in action. The source centroid is marked with the cross symbol. \emph{Right:} This radio source is yet not detected in 1.4 GHz, suggesting a steep radio power law index, which is likely due to the aging of a jet. The ellipse at the lower left corner of each panel represents the ASKAP beam size. Locations of the X-ray AGN neighbor \emph{ID2} are also indicated in both panels.  \label{fig:askap_contours} }
\end{figure*}

As shown in the left panel of Fig.~\ref{fig:askap_contours}, the Red Potato galaxy has a radio counterpart detected in 0.8 GHz, the center of which overlaps with the location of the galaxy, indicating a certain form of radio feedback in action. The radio source has a measured flux of $1.56 \pm 0.47$ mJy, the brightest of the large-scale structure. No significant detection is found in the 1.4 GHz map 2-$\sigma$ limit of 0.36 mJy; right panel) which, along with the 0.8 GHz detection, indicates a steep radio spectral index (Sect.\ \ref{subsec:analysis_xray_radio}).  On the 0.8 GHz map, the radio source appears elongated in a direction slightly offset from north-south, compared to the shape of the ASKAP beam shown in the lower left corner of the left panel. Because the full beam size of the ASKAP map is as large as 20 arcsec, it remains difficult to examine the radio morphology. Higher resolution follow-up radio observations are needed to pinpoint the origin of the radio emission.

\end{appendix}

\end{document}